\documentclass[prb,twocolumn,longbibliography,superscriptaddress,preprintnumbers,floatfix]{revtex4-2}
\usepackage[colorlinks,bookmarks=true,citecolor=blue,linkcolor=blue,urlcolor=blue]{hyperref}
\usepackage{dcolumn,graphicx,amsfonts,amsthm,bm,color,appendix,float}

\usepackage{braket}
\usepackage[normalem]{ulem}
\usepackage[version=4]{mhchem}
\usepackage{siunitx}
\DeclareSIUnit{\angstrom}{\mbox{\normalfont\AA}}
\usepackage{amsmath}
\usepackage{amssymb}
\usepackage{CJKutf8}
\usepackage{mathtools}
\usepackage{enumitem}
\usepackage{booktabs}
\usepackage{bbold}

\usepackage[dvipsnames]{xcolor}
\definecolor{pal0}{rgb}{0.8941, 0.102 , 0.1098}
\definecolor{pal1}{rgb}{0.2157, 0.4941, 0.7216}
\definecolor{pal2}{rgb}{0.302 , 0.6863, 0.2902}
\definecolor{pal3}{rgb}{0.5961, 0.3059, 0.6392}
\definecolor{pal4}{rgb}{1.    , 0.498 , 0.    }

\usepackage{tikz}
\usetikzlibrary{arrows, automata}
\usetikzlibrary{decorations.pathreplacing,decorations.markings}
\usetikzlibrary{decorations.pathmorphing}
\usetikzlibrary{shapes.geometric}

\tikzset{
    dot/.style = {fill=black, circle, inner sep=2.5},
	circ/.style={circle, draw, minimum size=2em,inner sep=1},
	sq/.style={rectangle, draw, minimum size=2*(#1-\pgflinewidth), inner sep=5pt, outer sep=0pt,fill=white},
	tedge/.style={ultra thick},
	on each segment/.style={
    decorate,
    decoration={
      show path construction,
      moveto code={},
      lineto code={
        \path [#1]
        (\tikzinputsegmentfirst) -- (\tikzinputsegmentlast);
      },
      curveto code={
        \path [#1] (\tikzinputsegmentfirst)
        .. controls
        (\tikzinputsegmentsupporta) and (\tikzinputsegmentsupportb)
        ..
        (\tikzinputsegmentlast);
      },
      closepath code={
        \path [#1]
        (\tikzinputsegmentfirst) -- (\tikzinputsegmentlast);
      },
    },
  },
  mid arrow/.style={postaction={decorate,decoration={
        markings,
        mark=at position .5 with {\arrow[#1]{stealth}}
      }}},
  snake/.style={decorate, decoration=snake}
}

\newcommand{\n}[1]{\left| #1 \right|}

\renewcommand{\v}[1]{\boldsymbol{#1}}

\usepackage{dsfont}
\usepackage{bbold}

\renewcommand{\AA}{\widehat{A}}

\renewcommand{\L}{\mathcal{L}}

\newcommand{\tr}{\operatorname{Tr}}

\newcommand{\e}{{e}}

\renewcommand{\i}{i}
\newcommand{\dd}{{d}}

\begin{document}

\title{
Phonons in Electron Crystals with Berry Curvature
}

\author{Junkai Dong (\begin{CJK*}{UTF8}{bsmi}董焌\end{CJK*}\begin{CJK*}{UTF8}{gbsn}锴\end{CJK*})}
\thanks{These authors contributed equally.}
\affiliation{Department of Physics, Harvard University, Cambridge, MA 02138, USA}

\author{Ophelia Evelyn Sommer}
\thanks{These authors contributed equally.}
\affiliation{Department of Physics, Harvard University, Cambridge, MA 02138, USA}

\author{Tomohiro Soejima (\begin{CJK*}{UTF8}{bsmi}副島智大\end{CJK*})}
\affiliation{Department of Physics, Harvard University, Cambridge, MA 02138, USA}

\author{Daniel E. Parker}
\affiliation{Department of Physics, University of California at San Diego, La Jolla, California 92093, USA}

\author{Ashvin Vishwanath}
\email{avishwanath@g.harvard.edu}
\affiliation{Department of Physics, Harvard University, Cambridge, MA 02138, USA}

\begin{abstract}
Recent advances in 2D materials featuring nonzero Berry curvature have inspired extensions of the Wigner crystallization paradigm. This paper derives a low-energy effective theory for such quantum crystals, including the anomalous Hall crystal (AHC) with nonzero Chern number. First we show that the low frequency dispersion of phonons in AHC, despite the presence of Berry curvature, resembles that of the zero field (rather than finite magnetic field) Wigner crystal due to the commutation of translation generators. We explain how key parameters of the phonon theory such as elastic constants and effective mass can be extracted from microscopic models, and apply them to two families of models: the recently introduced $\lambda$-jellium model and a model of rhombohedral multilayer graphene (RMG). In the $\lambda$-jellium model, we explore the energy landscape as crystal geometry shifts, revealing that AHC can become `soft' under certain conditions. 
This causes transitions in lattice geometry, although the quantized Hall response remains unchanged.
Surprisingly, the Berry curvature  seems to enhance the effective mass, leading to a reduction in phonon speed. For the AHC in RMG,  we obtain estimates of phonon speed and shear stiffness. We also identify a previously overlooked `kineo-elastic' term in the phonon effective action that is present in the symmetry setting of RMG, and leads to dramatic differences in phonon speeds in opposite directions. 
We numerically confirm these predictions of the effective actions by time-dependent Hartree-Fock calculations.
\end{abstract}

\maketitle

\section{Introduction}

The topic of electron crystallization has a rich history, originating with Wigner's analysis~\cite{wigner_interaction_1934}. Since then, research has expanded to include the crystallization of electrons in strong magnetic fields and on the surface of liquid helium~\cite{lozovik_crystallization_1975,grimes_evidence_1979,andrei_observation_1988,santos_observation_1992}.  Driven by significant advances in the synthesis and tunable doping of two-dimensional (2D) materials, there has been a surge of renewed interest in this problem~\cite{yoon_wigner_1999,ma_thermal_2020,smolenski_signatures_2021,zhou_bilayer_2021,yang_experimental_2021,falson_competing_2022,sung_observation_2023,tsui_direct_2023,xiang_quantum_2024}. Of particular interest are several cases involving valley degrees of freedom like graphene and TMD heterostructures, where spontaneous valley polarization leads to broken time reversal symmetry and parent bands with Berry curvature~\cite{GrapheneRMP,TMDNRM}. A low density of electrons doped into these parent bands sets the stage for electron crystallization in the presence of inherent Berry curvature~\cite{AHC_Senthil,AHC_Yahui,AHC1}. An intriguing possibility is the emergence of a Chern insulator upon spontaneous breaking of translation symmetry, dubbed the anomalous Hall crystal (AHC)~\cite{AHC_Yahui,AHC1}, which may be thought of as a zero-field counterpart of the Hall crystal proposed in Refs.~\cite{kivelson_cooperative_1986,halperin_compatibility_1986,kivelson_cooperative_1987,tesanovic_hall_1989}. The intriguing observation of quantum anomalous Hall effects in rhombohedral 4-5-6-7 layer graphene~\cite{lu2024fractional, waters2025chern, xie2025tunablefractionalcherninsulators, choi2024electricfieldcontrolsuperconductivity, Lu_2025, xiang2025continuouslytunableanomaloushall} under a strong displacement field and a weak moiré potential strongly motivates further study of the AHC including its stability~\cite{guo2024fractional, kwan2023moir, AHC2, Zhihuan_Stability, yu2024moir,zhou2024newclassesquantumanomalous, crepel2025efficient, bernevig2025berrytrashcanmodelinteracting,Zeng_sublattice_structure, Tan_parent_berry, Tan_FAHC, Zeng_sliding}. 

Despite the growing interest in electronic crystals,  the formulation of an effective theory of low energy phonons in the most general setting remains an open challenge. In this work, we derive the general form of the low-energy phonon action for electronic crystals, and outline a calculational scheme to obtain the parameters of this theory from a microscopic model. Our focus will be on crystals that are insulating once they are pinned, encompassing both anomalous Hall and traditional Wigner crystals. We work within a general framework with minimal symmetry constraints, allowing for the possible absence of time-reversal and  rotational symmetries, in addition to breaking Galilean invariance. Surprisingly, we find that certain terms in the effective theory, which initially seem permissible based on symmetry considerations, are actually forbidden. Conversely, we identify a kineo-elastic term that intriguingly couples strain with the time derivative of the displacement vector, which to our knowledge had not been pointed out in previous studies.

We derive a general formalism that enables us to immediately address a fundamental question: What is the low-energy dispersion of phonons in the anomalous Hall crystal (AHC)? It is well known that the phonon dispersions of a regular Wigner crystal differ considerably depending on whether a magnetic field is present~\cite{Fukuyama_magnetic_1975,Bonsall-Maradudin}. While at first glance the AHC seems to share many properties in common with the Wigner crystal in a magnetic field, including broken time reversal symmetry, we find surprisingly that the low energy phonon dispersion actually follows that of the zero field Wigner crystal. This conclusion rests  purely on the fact that translations in orthogonal directions commute and mirrors  arguments for the counting of Goldstone modes \cite{Nambu,Goldstone,schafer_kaon_2001, nambu_spontaneous_2004, watanabe_unified_2012, watanabe_number_2011, leutwyler_nonrelativistic_1994, brauner_goldstone_2005, brauner_goldstone_2007, watanabe_counting_2020}. This highlights the delicate interplay between time reversal symmetry breaking, translation symmetry and magnetic fields. 

We extract the phonon action coefficients by matching the phonon response with the long wavelength response of the microscopic electronic action. Due to the sensitivity of the ground state to the boundary conditions, the response can be calculated by considering a translational gauge twist, i.e. by modifying the periodic boundary conditions by a translation.
It is worth briefly describing the key steps in this approach. A useful analogy is  with {\em superfluid} effective actions, where U(1) charge symmetry is spontaneously broken and the relevant parameters governing the effective theory are superfluid stiffness and charge compressibility. In crystals, where translations are broken  the analogous   quantities we need to calculate are the elastic moduli and the effective mass.
We bridge the gap to microscopic models by providing a method for determining the parameters of the effective theory in terms of the intrinsic properties of microscopic electronic models, which can be calculated with any method that provides an estimation on ground-state energies.

In this work, we will work with the self-consistent Hartree-Fock (SCHF) approximation. For the elastic moduli, we consider the variation of the SCHF energy upon deforming the unit cell of the crystal. For the effective mass, we extract the `momentum compressibility' by evaluating the change in the ground state energy in the presence of a boost transformation, {which applies a chemical potential for the broken symmetry charge, i.e. momentum}. In the absence of Galilean invariance this is no longer just the electron mass, and is a key characteristic of the crystalline state.  The mixed kineo-elastic term is obtained by combining these procedures.

We then apply our procedure to $\lambda$-jellium, a model with tunable Berry curvature that encompasses both traditional Wigner crystals and anomalous Hall crystals, as well as models of rhombohedral multilayer graphene. A useful check on our calculations is provided by the time-dependent Hartree-Fock (TDHF) calculations that we present, which agree well at low energies with our phonon effective theory with no fitting parameters. 

Let us briefly summarize the results of the numerical investigation.
In the $\lambda$-jellium model, we find that the addition of Berry curvature tends to make the crystals softer and heavier, particularly when their Berry curvature is not too concentrated.
In some parts of the phase diagram, the crystal becomes soft enough to destabilize the triangular lattice. We compute the full energy landscape at those points to find  square or rhombic crystals as global minima.

The subtlety of the crystal lattice shape carries over to the rhombohedral multilayer graphene, so we focus on a parameter regime where the triangular lattice is stabilized. By applying our procedure, we find that the kineo-elastic term is non-zero and large: the velocities in the $+x$ and $-x$ directions differ by as much as $20\%$. Furthermore the transition temperature can be estimated  from the stiffness using the KTHNY theory of melting to be $1.9\si{\kelvin}$, showing that the crystal may be stable at experimentally relevant temperatures~\cite{KTMelting_A,KTMelting_B,NelsonKosterlitz,NelsonMelting,FisherShear}.

We now review the relationship of this work to previous works. Wigner crystallization in a band with Berry curvature was previously studied in Ref.~\cite{joy2023wignercrystallizationbernalbilayer}.
In deriving the phonon effective action, our approach is similar to obtaining the effective action by gauging the broken symmetry, which has been considered widely for internal symmetries, see \cite{LEUTWYLER1994165,braunerEffectiveFieldTheory2024} and references therein, and in the context of spacetime symmetries in~\cite{hidakaEffectiveFieldTheory2015}.
In the context of unidirectional ordering in superconductors, a similar approach, taking only the pairing field into account appeared in \cite{samokhinGoldstoneModesLarkinOvchinnikovFuldeFerrell2010}.

The instability of triangular crystals has been studied previously. In the context of Wigner crystals, Ref.~\cite{HF_Wigner_crystal} found a square lattice antiferromagnetic   crystal in Hartree-Fock theory. 
Meanwhile, in the context of AHC, its instability to lattice deformation has been explored using time-dependent Hartree-Fock~\cite{kwan2023moir} and elasticity theory~\cite{desrochers2025elasticresponseinstabilitiesanomalous}. An instability to expanding unit cell~\cite{zhou2024newclassesquantumanomalous, waters2024topological} has also been explored.
Moreover, the purely kinetic part of the effective action and some additional terms were extracted by~\cite{Zeng_sliding}.

In contrast, our analysis considers \textit{all} symmetry allowed terms, including the previously-overlooked kineo-elastic term. We thus present the first full low-energy elastic theory of electron crystals.  Furthermore, we provide simple methods for computing \textit{all} coefficients in the phonon effective action from microscopic models, validated by a rigorous comparison with TDHF numerics. This provides a simple and easily computed recipe to construct a quantitatively accurate low energy phonon theory of any electronic crystal.
   
The rest of the paper is organized as follows. In Sec.~\ref{sec:effective_field_theory}, we review the elastic theory for phonons. 
We then develop a formalism for computing the effective action from ``gauge twists'', which change the boundary condition of the system. This gives rise to a straightforward numerical algorithm for computing all coefficients of the effective theory for phonons. In Sec.~\ref{sec:numerical}, we apply this method to the $\lambda$-jellium model and rhombohedral multilayer graphene. We benchmark our results against TDHF calculations and find excellent agreement with the low frequency spectrum. Armed with these tools, we take a detailed tour of the phase diagram of $\lambda$-jellium model, and compute its full energy landscape versus the crystal lattice shapes. We apply a similar method to rhombohedral pentalayer graphene, finding a locally stable anomalous Hall crystal with large kineo-elastic coupling. We close with Sec.~\ref{sec:discussion}, where we discuss future directions.

\section{Effective Field theory of Phonon Modes in Quantum Crystals}
\label{sec:effective_field_theory}

In this section, we derive the method for computing the coefficients of the low-energy effective field theory. Our main technique is to couple the system to background \textit{translation} gauge twist.

\subsection{The elastic theory for phonons}
\label{sec:elastic_theory}
To set the stage, let us review the elastic theory of phonons, while taking care not to drop any symmetry-allowed terms. We consider an electronic crystal whose equilibrium electron locations are given by $\v{R}_i \in \Lambda$, where $\Lambda$ is the crystalline lattice. We assume that all electrons crystallize, so that in the presence of weak pinning we have an insulator. We take the average density of electrons to be $n$. We define the {\em displacement field} $\v u(\v{R}_i)$ such that $\v u(\v{R}_i) + \v{R}_i$ is the location of the electron originally at $\v{R}_i$. By taking the long-wavelength limit, we convert $\v u$ to a continuous field and analyze its action.

In the presence of Coulomb interaction, it is convenient to split the action into the local part and the Coulomb part: 

\begin{equation}
    S^\mathrm{phonon}[\v{u}] = S^\mathrm{local}[\v{u}] + S^\mathrm{Coulomb}[\v{u}]
\end{equation}

 We will impose translation symmetry, implemented by $\v{u}\rightarrow \v{u}+\v\epsilon $ for arbitrary constant $\v\epsilon$. Crucially, we do not require either time-reversal symmetry or other lattice symmetries such as inversion. The local action is obtained by keeping all symmetry allowed terms up to second order in derivatives and quadratic in the displacement fields:
\begin{align}
S^\mathrm{local}[\v u] & =\rho \int \dd t\dd \v r\,\left[\frac{1}{2}m_{ab}\dot{u}_a\dot{u}_b-\frac{1}{2} \lambda_{abcd}u_{ab}u_{cd}\right.\nonumber \\
& \left.\qquad\qquad\qquad+\frac{1}{2}\beta \epsilon_{ab} u_a\dot{ u}_b-\ell_{abc}\dot{ u}_a\partial_bu_c\right],
\label{eq:phonon_action}
\end{align}
Here, we introduced $u_{ab} = \frac{1}{2}[\partial_a u_b + \partial_b u_a]$ for convenience and $\epsilon_{ab}$ is the two dimensional antisymmetric tensor. 
The equation of motion derived from the phonon action is invariant under a constant shift in $\v u$, as required by translation symmetry. Let us now go through the coefficients one by one: 1) $m_{ab}$ is the effective mass that controls the kinetic energy of sliding crystal. 2) $\lambda_{abcd}$ is the stiffness tensor, which measures the energy cost associated with deforming the crystal. 3) $\beta$ is the Berry phase term, which generates a Lorentz force on the center of mass motion of the crystal. 4) $\ell_{abc}$ is a \textit{kineo-elastic} term, which couples the strain on the crystal with its velocity.

We note that the Berry phase term is at first sight not invariant under translation. However, upon shifting by $\Delta u$ it generates a term of the form $\beta \epsilon_{ab} \Delta u_a \dot{u}_b/2$. This is a total derivative and does not affect the equation of motion. Time-reversal symmetry forbids this term. We consider Hamiltonians that break time-reversal, however, so this term is \textit{a priori} allowed in the action.

The Coulomb term, on the other hand, is given by
\begin{align}
S^\mathrm{Coulomb}[\v{u}]
= \rho^2e^2 \int dt d\v{r} d\v{r}' \frac{1}{\epsilon |\v{r} - \v{r}'|} u_{aa}(\v{r}) u_{bb}(\v{r}')\label{eq:coulomb}
\end{align}
We will set the electron charge to be $e=1$ subsequently.

As an example, we will now impose $C_3$ symmetry, which is respected by all models we consider in this paper. The $C_3$ symmetry will constrain the parameters in the effective field theory. The kinetic term becomes isotropic: $m_{ab} = m\delta_{ab}$. The elastic term also simplifies greatly to be parameterized by two parameters:

\begin{equation}
\label{eq:C3sym_elastic}
    \frac{1}{2}\lambda_{abcd}u_{ab}u_{cd} = \frac{K}{2}(u_{xx}+u_{yy})^2 + \frac{\mu}{2}[(u_{xx}-u_{yy})^2+(2u_{xy})^2]
\end{equation}
in which $K$ characterizes a response to the expansion of volume, and $\mu$ characterizes the response to volume-preserving deformations, such as strains. This is also why we will refer to this term as ``shear stiffness'' below. Surprisingly, all of these terms behave as if the system is isotropic: they are all invariant under continuous rotations.

The kineo-elastic term deserves more attention. The presence of this term changes the canonical momentum in the presence of strain to
\begin{equation}
    \pi_a = \frac{\partial L}{\partial {\dot{u}_a}} = m {\dot{u}_a} - \ell_{abc}\partial_b u_c.
\end{equation}
For a fixed value of $\dot{\v{u}}$ and $\partial_a \v{u}$, however, the Hamiltonian is fully determined by the effective mass and stiffness, and does not depend on the kineo-elastic term.

To further analyze this term, we note that $\ell_{abc}$ carries three vector indices. Thus, it will carry angular momentum either $3$ or $1$; the $C_3$ symmetry will constrain such that only the angular momentum $3$ terms remain. Those can be parameterized by two real numbers:

\begin{equation}
\label{eq:C3sym_kineoelastic}
\begin{aligned}
    \ell_{abc}u_a\partial_b u_c &= \ell(\dot{u}_{x}(u_{xx}-u_{yy})-\dot{u}_y(2u_{xy})) \\&+ \ell'(\dot{u}_{y}(u_{xx}-u_{yy})+\dot{u}_x(2u_{xy})).
\end{aligned}
\end{equation}

We can always choose to rotate the system such that $\ell'=0$. Since this leaves the rest of the Lagrangian invariant, we will only consider the case where $\ell\neq 0$. We will later prove $\beta=0$ for crystals of interest. Assuming this, we can solve for the poles of the Green's function for the phonons. We then find that the leading order dispersion in the $\v q\rightarrow 0$ limit of the longitudinal and transverse phonons are:

\begin{equation}
\begin{aligned}
    \omega_l(\v q) &= \sqrt{\frac{2\pi \rho|\v q|}{\epsilon m}},\quad
    \omega_t(\v q) = v_t(\hat{\v q})|\v q|,\\
    v_t(\hat{\v q}) &= \frac{\cos(3\theta_{\v q})\ell+\sqrt{4m\mu+\cos^2(3\theta_{\v q})\ell^2}}{2m}
\end{aligned}
\label{eq:phonon_spectrum_and_velocity}
\end{equation}
in which $\theta_{\v q}$ is the angle between the $x$ axis and $\v q$. We note that while the longitudinal phonon has a dispersion $\propto \sqrt{q}$, the transverse phonons have a finite velocity that varies in direction. Notably, as a result of the kineo-elastic term, the forward and backward speed of  phonons is in general different and an angle dependent quantity, where $v_t(\hat{\v q})-v_t(-\hat{\v q})= \cos(3\theta_{\v q})\ell/{m}$.
We remark that the presence of the kineo-elastic term is purely a consequence of the low symmetry of the system, and is generically present for any crystal. 

We find that $K$ does not appear in the leading order contributions to the phonons; its contributions are one order higher in $|\v q|$ compared to the Coulomb term $S^{\textrm{Coulomb}}[\v u]$. Thus, we drop the consideration for $K$ whenever we discuss $C_3$ symmetric crystals.

\subsection{Low-energy action from the translational gauge twist}
Recall, in deriving an effective low energy theory for superfluids it is helpful to consider the system with periodic boundary conditions, that we then twist by phase rotation. The superfluid density can be extracted from the energy cost of the twist, at least when the only low energy modes are Nambu-Goldstone modes. As an analogous procedure here, we will twist boundary conditions by the spontaneously broken symmetry i.e. translations, and extract the energy cost to obtain the stiffness.  This procedure is similar to obtaining the effective goldstone theory by gauging the symmetry, and appealing to the Anderson-Higgs mechanism, see for example \cite{braunerEffectiveFieldTheory2024} and references therein, with the notable advantage that we don't need to consider the effect of the curvature of the gauge field, which is a significant practical simplification.

A translational
gauge twist is a time and spatially dependent translation  $\v r\to \v r'=\v r-{\v \theta}(\v r',t)$. We will only need gauge twists corresponding to constant gradients. In analogy with the superfluid, we consider a system on a torus with a fixed periodic boundary condition. This gauge twist may be obtained by acting with a unitary that takes the form $U_{\v\theta} = \exp(-i\int d\v r\, \v \pi(\v r) \cdot \v \theta)$, where $\v\pi$ is the momentum density, which is the charge density of the translation symmetry.
Let $H$ be the Hamiltonian obtained in the absence of gauge twist. By considering the time evolution of a state $U_{\v\theta}\ket{\psi}$ as being governed by $H$, the time evolution of $\ket{\psi}$ is governed by the twisted Hamiltonian
\begin{equation}
H[{\v \theta}]=U_{\v\theta}^\dagger HU_{\v\theta}-\i U_{\v\theta}^\dagger\dot{U}_{\v\theta}
\label{eq:twisted_general_ham},
\end{equation}
which we take to act of a Hilbert space with fixed $\v\theta$ independent boundary conditions. 
Its precise form depends on the choice of the microscopic Hamiltonian, and we provide the form for fermionic Hamiltonians in Sec.~\ref{sec:application_to_fermionic_hamiltonian}.

While this is a unitary transformation in the infinite plane, it relates sectors with different choices for periodic boundary conditions. Therefore, with a fixed boundary condition, the spectrum changes before and after the transformation. With this choice of Hamiltonian   Eq.~\ref{eq:twisted_general_ham}, we can write the effective action as

\begin{equation}
	\e^{\i \Gamma[{\v \theta}]}=\tr \left[T\e^{-\i\int_{t_i}^{t_f} H[{\v \theta}]\dd t}\right]
\end{equation}
where $T$ is the time ordering symbol, and by trace we really mean inserting
appropriate boundary conditions at the initial and final times corresponding to the ground state satisfying ${\v \theta}$ independent boundary conditions.

$\Gamma[\v \theta]$ contains all dynamical information coming from low-energy modes of the system that couple to $\v{\theta}$. By understanding long-time long-wavelength behavior of $\Gamma[\v \theta]$, we can extract the effective theory of the system.

\subsection{Matching with the phonon theory}
Let us now suppose there exists a phonon field $\v u$ whose action characterizes the low energy response. Let us also assume that $\v{u}$ transforms under the translational gauge twist as $\v u \to \v u-{\v \theta}$.
Therefore the action is modified by the gauge twist, becoming:
\begin{equation}
    S^{\mathrm{phonon}}[\v{u}] \to S^{\mathrm{phonon}}[\v{u} - \v{\theta}].
\end{equation}

The effective action for $\v \theta$ arises from integrating phonon modes
\begin{equation}
	\e^{\i\Gamma[{\v \theta}]}=\int\mathcal{D}\v u\,\e^{\i S^\mathrm{phonon}[\v u-{\v \theta}]},
\end{equation}
where the integration $\mathcal{D}\v{u}$ is only over boundary-condition-respecting $\v{u}$. Now consider gauge twisting by $\v \theta$ which obeys different boundary conditions 
\begin{equation}
\v\theta_i(\v r)=\v\theta(\v r,t_i),\quad \v \theta_f(\v r)=\v\theta(\v r,t_f).
\end{equation}
If $\v \theta$ is chosen to satisfy the equations of motions in the bulk, then we may consider expanding the action to quadratic order in $\v u$, suppressing space-time indices for brevity
\begin{align}
    S^\mathrm{phonon}[\v u-\v \theta]&=S^\mathrm{phonon}[-\v \theta]+\int\frac{\delta S[-\v\theta]}{\delta\theta^a}u^a\\&+\frac{1}{2}\int\frac{\delta^2 S[-\v\theta]}{\delta\theta^a\delta \theta^b}u^a u^b\\
    &=S^\mathrm{phonon}[-\v \theta]+S^\mathrm{phonon}[\v u]
\end{align}
The linear term vanishes since $\v \theta$ obeys the bulk equation of motion and the quadratic $\v u$ term must be precisely the original action evaluated at $\v u$ since the action is {\em quadratic}. Since the phonon no longer couples to $\v \theta$, we can trivially perform the path integral, and noting that $S[-\v \theta]=S[\v \theta]$ we see that up to an irrelevant additive constant
\begin{equation}
    \Gamma[\v{\theta}] = S^{\mathrm{phonon}}[\v\theta].
\end{equation}

Therefore, the effective action in terms of an on-shell configuration of $\v \theta$ contains all information about the effective action $S^{\mathrm{phonon}}[\v{u}]$.

\subsection{Extraction of the coefficients}
\label{sec:extraction}
Let us first consider the Berry phase term $\beta$ in the phonon action. Here we show that the coefficient of this Berry phase term is constrained such that
\begin{equation}
    \beta = 0, \quad\textrm{when } [\hat{p}_x,\hat{p}_y]=0
    \label{Eq:beta-commutator}
\end{equation}
where $\hat{\v p} = \sum_{i}\hat{\v p}_{i}$ is the total momentum operator for all the particles labeled by $i$, and $N$ is the number of particles.
This formula echoes the argument for counting of Nambu-Goldstone bosons in nonrelativistic systems~\cite{Nambu,Goldstone,schafer_kaon_2001, nambu_spontaneous_2004, watanabe_unified_2012, watanabe_number_2011, leutwyler_nonrelativistic_1994, brauner_goldstone_2005, brauner_goldstone_2007, watanabe_counting_2020}.

In the models we consider the momentum operators $\hat{p}_x,\,\hat{p}_y$ {\em commute}, regardless of whether we have a regular Wigner crystal or an AHC. Therefore we find that $\beta=0$ and this term is {\em absent} in these models. 
Conversely, systems with non-commuting translations, such as charged particles in a magnetic field, can pick up nonzero Berry phase term $\beta$. It is well known that the dispersion relations for Wigner crystal phonons are strongly modified by a magnetic field, due to the onset of this term~\cite{Fukuyama_magnetic_1975,Bonsall-Maradudin}. Important subtleties can arise when a magnetic field is present, which we will delay to a subsequent work~\cite{AHC5}.

Let us now derive   Eq.~\ref{Eq:beta-commutator}. We extract $\beta$ by considering the boost response in a system that has been displaced a constant amount $\Delta x$ in an orthogonal direction. If $\beta\neq0$ the on-shell motion is generically that of the cyclotron orbits, but by taking the time interval $t_f-t_i$ to be much shorter than the cyclotron period, the differences between this and rectilinear motion in the intervening time are irrelevant for extracting $\beta$. From the representation of the effective action in terms of phonons, this response depends only on $\beta$:
\begin{align}
    \frac{\partial}{\partial \dot{\theta}_y}\Gamma[\theta_x=\Delta x,\theta_y=\dot{\theta}_y(t-t_i) ] = \frac{1}{2}\int \dd t N\beta\Delta x.
\end{align}
By comparison, we may also evaluate the response from the Hamiltonian $H[\v \theta]=U_{\v \theta}^\dagger HU_{\v \theta}-\i U^\dagger_{\v \theta}\dot{U}_{\v \theta}$. Since the Hamiltonian is invariant under uniform translations, the only change in the action must arise from the change in the Maurer-Cartan form. A time-dependent uniform translation is generally implemented by the unitary $U_{\v \theta}=e^{-\i \v \theta\cdot \hat{\v p}}$. As such, the Maurer-Cartan form to quadratic order in $\v \theta$ is 
\begin{equation}
    -\i U^\dagger_{\v \theta}\dot{U}_{\v \theta}=-\dot{\theta}_y\hat{p}_y+\dots
\end{equation}
in which we have used $[\hat{p}_x,\hat{p}_y]=0$. Thus
\begin{align}
    \frac{\partial}{\partial \dot{\theta}_y}\Gamma&= -\frac{\partial}{\partial \dot{\theta}_y}\int\dd t \braket{H[\v\theta]}=\int\dd t\braket{\hat{p}_y}
\end{align}
where expectation values are taken in the translated but stationary ground state. Clearly when the momenta commute $\braket{\hat{p}_y}=0$, from which equation Eq.~\ref{Eq:beta-commutator} immediately follows.

We now move on to extract other coefficients of the effective action.
When the momenta commute so that $\beta=0$, the effective action only depends on gradients of $\v\theta$. Then $\v{\theta}$ with arbitrary constant spatial and temporal gradients are guaranteed to satisfy the equations of motion. 
In such cases, the Maurer-Cartan form $-i U_\theta^\dagger \dot{U}_\theta = -\dot{\theta}_x \hat{p}_x - \dot{\theta}_y \hat{p}_y$ is also time independent. Therefore, $H[\v{\theta}]$ is time-independent, and we can write the action as:

\begin{equation}
    \e^{\i \Gamma[{\v \theta}]} = 
    \e^{-i (t_f - t_i) \langle H[\v\theta] \rangle},
\end{equation}
where $\langle H[\v \theta] \rangle$ is the ground state energy of $H[\v \theta]$ in the original, fixed boundary condition.
Therefore we get
\begin{equation}
    \Gamma[{\v \theta}]=-(t_f-t_i)\,\braket{H[{\v \theta}]}
\end{equation}
Thus expanding the ground state energy to quadratic order and taking derivatives, we get
\begin{align}
    \label{eq:mass_extraction} m_{ab} &=-\frac{1}{N}\left.\frac{\partial^2}{\partial\dot{\theta}_a\partial \dot{\theta}_b}\braket{H[{\v \theta}]}\right|_{\v \theta=0}\\ 
    \label{eq:ell_extraction} \ell_{abc}&=\frac{1}{N}\left.\frac{\partial^2}{\partial (\partial_{b}\theta_c)\partial \dot{{\theta}}_a }\braket{H[{\v \theta}]}\right|_{\v \theta=\v 0}\\ 
   \label{eq:lambda_extraction}\lambda_{abcd} &=\frac{1}{N}\left.\frac{\partial^2}{\partial \theta_{ab}\partial \theta_{cd}}\braket{H[\v \theta]}\right|_{\v \theta=\v 0}
\end{align}
where $N$ is the total number of particles, and we introduced $\theta_{ab} = \frac{1}{2}(\partial_a \theta_b + \partial_b \theta_a)$. We emphasize again that these expectation values are taken with respect to the ground state of $H[{\v \theta}]$, subject to boundary conditions that are independent of ${\v \theta}$.  The details of the extraction procedure, including convenient choices for  $\v{\theta}$ are given in Sec.~\ref{sec:numerical_implementation}.  

Although we will compute these expectation values within Hartree-Fock theory, it is important to note that the formalism developed here is applicable to any other numerical method for computing the ground state energy. 

\subsection{Application of the general procedure to electronic Hamiltonians}
\label{sec:application_to_fermionic_hamiltonian}
We now consider how to implement the previously described procedure for electronic systems.
Consider a two-dimensional translationally invariant electron gas with first quantized Hamiltonian of $N$ particles
\begin{align}
    H&=\sum_{i=1}^Nh (\hat{\v p}_{i})+\frac{1}{2}\sum_{i\neq j}V(\hat{\v r}_i-\hat{\v r}_j)
\end{align}
where $h$ is the bare Hamiltonian of the bands, which generally has internal state structure,
and $V$ is the interaction. We consider a system on a finite torus geometry with periods $\v L_{1,2}$. The gauge twist labeled by $\v \theta$ physically represents deforming and boosting this torus. Clearly due to the long range nature of the Coulomb interaction, a deformation that does not preserve the area of the system, will result in a super-extensive change in the total energy. The corresponding response is universal, and given in the phonon theory by equation Eq.~\ref{eq:coulomb}. As such, we may focus on unimodular deformations, and boosts so that for any single-particle wavefunction $\psi(\v r)$ the deformation takes the form $U_{\v \theta}\psi(\v r)=\psi(\v r-\v \theta)$. 
For a uniform boost ${\v v}$ and a deformation $M_{ab}$, that is 
\begin{equation}
\label{eq:theta_parametrization}
    \theta_a(\v r,t)=v_at+ M^{-1}_{ab}r_b,
\end{equation} 
the operator factorizes 
\begin{equation}
    U_{\v \theta}=\e^{-\i{\v v}t\cdot \v p}\e^{-\i\frac{1}{2}M^{-1}_{ab}[\hat{p}_a\hat{r}_b+\hat{r}_{b}\hat{p}_a]}=U^\mathrm{boost}_{{\v \theta}}U^\mathrm{unimod}_{\v \theta}
\end{equation}
The corresponding Maurer-Cartan form is 
\begin{equation}
\label{eq:maurer-cartan}
    -\i U_{\v \theta}^\dagger\dot{U}_{\v \theta}=-\left(U_{\v \theta}^\mathrm{unimod}\right)^\dagger{\v v}\cdot\hat{\v p} \,U_{\v \theta}^\mathrm{unimod}
\end{equation}
Thus the gauged Hamiltonian takes the form
\begin{equation}
    H[\v \theta]=\left(U_{\v \theta}^\mathrm{unimod}\right)^\dagger\left(H-{\v v}\cdot\hat{\v p}\right)U_{\v \theta}^\mathrm{unimod}
\end{equation}
This Hamiltonian should be understood to act on states with fixed boundary conditions, which for a single particle wavefunction take the form $\psi(\v r+\v L_{1,2})=\psi(\v r)$. On the other hand, it is straightforward to see that for any wavefunction $\psi(\v r)$ satisfying the original boundary conditions, the wavefunction 
$\tilde{\psi}=U^\mathrm{unimod}_{\v \theta}\psi$ satisfies boundary conditions $\tilde{\psi}(\v r+(1+M)\v L)=\tilde{\psi}(\v r)$. The state $\tilde{\psi}$ now transforms under the Hamiltonian

\begin{equation}
    \tilde{H}[\v \theta]=
    H-{\v v}\cdot\hat{\v p}
    \label{eq:boosted_hamiltonian}
\end{equation}
Numerically we find it most convenient to work with this boosted Hamiltonian acting on states obeying the unimodularly changed boundary conditions.

Note, using Eqs.~\ref{eq:theta_parametrization},~\ref{eq:boosted_hamiltonian} and the G\"uttinger-Hellmann-Feynman theorem~\cite{Guttinger,Hellmann,Feynman}, Eqs.~\ref{eq:mass_extraction},~\ref{eq:ell_extraction} can equivalently be cast as:
\begin{align}
       m_{ab} =\frac{1}{N}\frac{\partial}{\partial\dot{\theta}_b}\braket{\hat{p}_a}, \quad
       \ell_{abc}=-\frac{1}{N}\frac{\partial}{\partial(\partial_b\theta_c)}\braket{\hat{p}_a}
       \label{eq:ell_extraction2}
\end{align}
We find that these formulas are much more stable numerically. We will now briefly discuss how we numerically implement these derivatives.

\subsection{Numerical implementation}
\label{sec:numerical_implementation}
\begin{figure}
    \centering
    \includegraphics[width=0.95\linewidth]{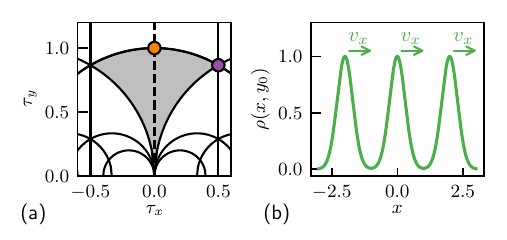}
    \caption{The numerical procedure for parameter extraction. 
    (a) Different unit cells of the same area are specified by the orientation $\phi$ and modular parameter $\tau$.
    All inequivalent choices of $\tau$ can be chosen by sampling a fundamental domain of the modular group (one choice is shaded in gray).  Solid lines represent the boundaries of the fundamental domain, which correspond to rhombic lattices. The dashed line corresponds to rectangular lattices. The square lattice (orange) corresponds to $\tau=i$ whereas the triangular lattice (purple) corresponds to $\tau=e^{2\pi i/3}$.
    (b) Boosting the system in the $x$ direction gives the crystal a center-of-mass velocity $v_x$ for its sliding motion.
    }
    \label{fig:methodology}
\end{figure}

In the discussion above, we have established that the coefficients of the phonon effective action can be extracted (1) by choosing different periods $\v L_{1,2}$ for the torus geometry and (2) by boosting the system with velocity $\v v$. We now discuss how each of these are numerically implemented.

\subsubsection{Different Torus Geometries}

To be concrete, let's consider implementing a non-zero shear strain on the system. This is equivalent to taking a pair of new lattice vectors

\begin{equation}
    \v L'_{a} = (I+M(\theta_{xy})) \v L_a,\quad M(\theta_{xy}) = \begin{bmatrix}
        0 & 2\theta_{xy} \\ 
        0 & 0
    \end{bmatrix}
\end{equation}

Thus, if we can extract the energy landscape of all possible torus geometries, we can find the shear stiffness $\mu$ easily. To do that, we need a good parametrization of the space of all torus geometries. Since all of our finite size calculations are performed at constant electronic densities and electron numbers, the lattice vectors are constrained that the area $|\v L_1\times \v L_2|$ remains unchanged. All such lattices are thus parametrized by $\phi$ the orientation of lattice vector $\v L_1$ and a modular parameter $\tau$ in the upper half complex plane:

\begin{equation}
    \tau = \frac{L_{2x}+iL_{2y}}{L_{1x}+iL_{1y}}
\end{equation}

Given $(\phi,\tau)$ and the area of the torus $A$, there is a unique pair of lattice vectors that correspond to these values:

\begin{equation}
    \v L_{1}=\sqrt{\frac{A}{\tau_y}}R(\phi)\begin{bmatrix}
        1\\
        0
    \end{bmatrix}, \quad \v L_{2}=\sqrt{\frac{A}{\tau_y}}R(\phi)\begin{bmatrix}
        \tau_x\\
        \tau_y
    \end{bmatrix}
\end{equation}
in which $R(\phi)$ corresponds to a rotation by $\phi$.

Different choices of the unit cell for the same torus correspond to a modular transformation  $
    \tau\to \frac{a\tau+b}{c\tau+d}$,
where $a,b,c,d\in\mathbb{Z}$, $ad-bc=1$. All the inequivalent values for $\tau$, corresponding to different choices of lattices, form the fundamental domain of the modular group, one of which is shown in Fig.~\ref{fig:methodology}(a). The triangular lattice corresponds to $\tau = e^{2\pi i /3}$, and the square lattice corresponds to $\tau=i$.  When the Hamiltonian possesses continuous rotational symmetry, the energy of the ground state cannot depend on $\phi$. Thus, the modular parameter $\tau$ is the only variable that determines the energy. In that case, we can choose $\phi=0$ without loss of generality. The strain $\theta_{xy}$ thus takes $\tau_x\to\tau_x+2\theta_{xy}\tau_y, \tau_y\to\tau_y$. The stiffness corresponding to the rest of volume-preserving deformations can be extracted in a similar way.

\subsubsection{Boost transformation and Galilean invariance}

Once the unimodular transformation is implemented, we boost the Hamiltonian according to  ~Eq.~\ref{eq:boosted_hamiltonian} (Fig.~\ref{fig:methodology}(b)), which we reproduce here for convenience:
\begin{equation}
    \hat{H}'(\v v) = \hat{H} - \v v\cdot\hat{\v p},
\end{equation}
where $\hat{\v p} = \sum_i \hat{\v p}_i$ is the center of mass momentum. 
We note the same boosting procedure was performed in Ref.~\cite{Zeng_sliding} for models of AHC and RMG.

With a concrete procedure for the boost operator, it is now easy to see why Galilean invariance fixes the effective mass. When the system is Galilean invariant, the many-body Hamiltonian of the system factorizes: 
\begin{equation}
    \hat{H} = \frac{\hat{\v p}^2}{2 N m_b} + \hat{H}_{rel}
\end{equation}
in which $\hat{H}_{rel}$ describes the relative motion of the electrons, whereas the first term describes the center of mass motion. $m_b$ is the bare mass of the electron.
The boost only couples to the center of mass motion.
Since the two terms decouple, the minimization of $\hat{H}_{rel}$ can be performed independently; its ground state energy will be denoted by $E_0$. Now we shall consider minimizing the Hamiltonian that describes the center of mass motion $\hat{\v p}^2/{2 N m_b} - \v v\cdot\hat{\v p}$: it is minimized when the state is an eigenstate of $\hat{\v p}$ with eigenvalue $N m_b \v v$, leading to $m=m_b$ by Eq.~\ref{eq:ell_extraction2}.

\subsubsection{Full extraction procedure for a $C_3$ symmetric crystal}
\label{sec:C3sym_extraction}

For a $C_3$ symmetric crystal, the elastic Lagrangian simplifies greatly. The effective mass tensor becomes isotropic: $m_{ab} = m\delta_{ab}$. From   Eq.~\ref{eq:C3sym_elastic}, the elastic tensor simplifies greatly to be only described by two parameters, $K$ and $\mu$. However, given that $K$ describes compression modes, its effects are dominated by the Coulomb interactions. From   Eq.~\ref{eq:C3sym_kineoelastic}, the kineo-elastic couplings also simplify to be described by two terms $\ell$ and $\ell'$. Thus, there are four numbers to be extracted in total: $m,\mu,\ell,\ell'$. For convenience, we give explicit formulas for their extraction below.

\begin{equation}
    \begin{aligned}
        m &= \frac{1}{N} \frac{\partial \braket{\hat{p}_x}}{\partial v_x}, \quad
        \mu = \frac{1}{4N}\frac{\partial^2 \braket{H[\theta_{xy}]}}{\partial \theta_{xy}^2},
        \\
        \ell &= \frac{1}{2N}\frac{\partial \braket{\hat{p}_y}}{\partial \theta_{xy}}, \quad
        \ell' = -\frac{1}{2N}\frac{\partial \braket{\hat{p}_x}}{\partial \theta_{xy}}.\\
    \end{aligned}
\label{eq:c3_extraction}
\end{equation}

\subsection{Electromagnetic response of anomalous Hall crystals}
The feature distinguishing an (anomalous) Hall crystal from an ordinary Wigner crystal is the electromagnetic response. 
At energies below the gap to electrons, the electromagnetic response is governed by bound charges and currents and a topological Chern-Simons term.
\begin{align}
	\Gamma^{E<E_\mathrm{pinning}}[A]&=\int \dd t\dd\v r\,\left[M B+\v P\cdot \v E\right]+\frac{C}{4\pi} \int A\dd A+\dots
\end{align}
corresponding to a Hall conductivity $\sigma_{xy}=C\frac{e^2}{h}$, magnetization density $M$ and polarization density $\v P$. In general there will also be higher order derivative terms.
However, phonons also enter the electromagnetic response by modifying the polarization and magnetization. 
The leading coupling can be understood by considering a uniform translation by $\v \theta$ which leaves the average $M$ unchanged, but shifts the polarization density $\v P\to \v P+e\rho \v \theta$ where $\rho$ is the average density, leading to the coupling: 
\begin{equation}
   S^{\mathrm{phonon-EM}}[\v u,A]=\int \dd t\dd \v r\,e\rho \,\v u\cdot \v E+\dots
\end{equation}
which can be directly verified by including electromagnetic potentials in the microscopic theory at the outset.

We can obtain the dominant response to external fields by first setting $\v k= 0$ and then integrating out the phonons. When $\beta = 0$, this leads to Drude response:
\begin{equation}
    \sigma_{ab}(\v{k} = \v 0,\omega\to 0)=\i\frac{e^2 \rho}{\omega+i0^+} \left [ m \right ]^{-1}_{ab}
\end{equation}
with the Drude weight given by
${\mathcal D}_{ab}=e^2\pi \rho \left [ m \right ]^{-1}_{ab}$. 

If $\beta\neq0$, the Drude weight necessarily vanishes. Curiously, since the Drude weight contribution to the conductivity goes as $\omega^{-1}$, the phonon contribution to both the static $\omega^{0}$ longitudinal \emph{and} Hall conductance is subleading.
To obtain subleading terms in the conductivity one needs higher-order time derivative terms in the phonon effective action which we do not retain here. We expect that in general the transverse conductivity is not quantized in the absence of pinning, since anomalous phonon velocity contributions destroy the quantization of the transverse conductivity, as was explicitly demonstrated in Ref.~\cite{Zeng_sliding}. 
However, if we introduce pinning, in the form of a mass term $\v u^2$ to the phonon action, then at low frequencies the Hall conductance becomes quantized to 
$\sigma_{ab}=\epsilon_{ab}\frac{C}{2\pi}$. We summarize this as follows:
\begin{equation}
    \sigma_{xy}(\v{k} = 0, \omega \to 0)
    = \begin{cases}
        \frac{C}{2\pi} & \text{pinned crystal.} \\
        \text{unquantized} & \text{unpinned crystal.}
    \end{cases}
\end{equation}

\section{Numerical Results}
\label{sec:numerical}
We now use the general formalism developed above to study the low-energy excitations of three increasingly-complex electronic crystals. First, we study the Wigner crystal phase of jellium, where we show our linear response theory reproduces standard results as a sanity check. Second, we showcase how our method works in the $\lambda$-jellium model with non-trivial Berry curvature, accurately reproducing the phonon velocity from far-more-expensive time-dependent Hartree-Fock calculations. The Berry curvature drives a transition to an anomalous Hall crystal, which we find to be \textit{heavier} and \textit{softer} than the jellium Wigner crystal --- to the extent that the softness destabilizes the hexagonal lattice in one regime. Performing a global stability analysis reveals its fate: it remains within the AHC phase but prefers a rhombic or square lattice. Third and finally we consider a microscopic model of rhombohedral pentalayer graphene. This model has neither time-reversal nor inversion symmetry, thus allowing the kineo-elastic term. At a point where a triangular AHC phase is stabilized, we find a giant sound velocity and a non-zero kineo-elastic term that gives rise to strong directional anisotropies in the speed of sound. 

\subsection{The $\lambda$-jellium model}
\label{sec:lambda_jellium}
\begin{figure}[htbp]
    \centering
    \includegraphics[width=\linewidth]{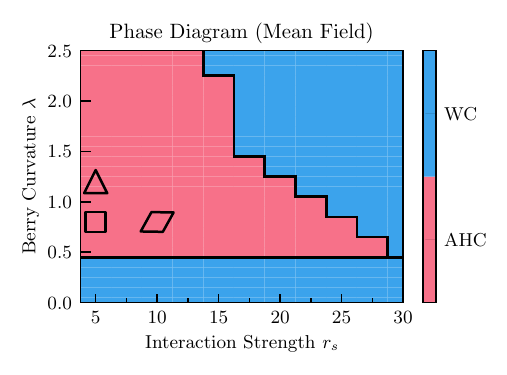}
    \caption{Hartree-Fock phase diagram of the $\lambda$-jellium model, which shows the competition between crystalline phases. Fermi liquid phases in the phase diagram are out of the parameter regimes shown here.
    Colors correspond to Wigner Crystals and Anomalous Hall Crystal. 
    The parameter combinations $(\lambda,r_s) = (0.8,5), (0.8,10), (1.2,5)$ will be studied in more careful detail in a later part of the manuscript; the anomalous Hall crystals take different shapes at those parameter points.
    All data are computed from SCHF ($18\times18$) with $13$ bands taken, where a triangular unit cell was assumed.
    }\label{fig:lambda_jellium_phase_diagram}
\end{figure}

\subsubsection{Hamiltonian and Mean-Field Phase Diagram}
\label{sec:lambda_jellium_EFT}

We now apply our linear response procedure to study the elastic theory of $\lambda$-jellium. $\lambda$-Jellium is a model, recently proposed by some of us, that extends the spinless jellium model with a parameter $\lambda$ that adds Berry curvature without modifying Coulomb interactions or the quadratic dispersion. When $\lambda  = 0$, the model reduces to normal jellium, whose Wigner crystal we study first as a sanity check.

The Hamiltonian of $\lambda$-Jellium is
\begin{equation}
	\hat{H}
    = \Delta \sum_{i=1}^N \begin{bmatrix}
        -\lambda^2 \nabla_i^2 & i\lambda \partial_i\\
        i\lambda \bar{\partial}_i & 1
    \end{bmatrix}
    - 
	\sum_{i=1}^N \hat{I}_2 \frac{\nabla_i^2}{r_s^2} + \frac{2}{r_s} \sum_{i<j}^N \frac{1}{\n{\v{r}_i - \v{r}_j}},
    \label{eq:lambda_Jellium_Hamiltonian}
\end{equation}
where $\Delta$ is taken to be large.

Following conventions of Ref.~\cite{tanatar1989ground}, length is measured in units of the interparticle distance $a$ and energy is measured in Rydbergs~(Ry), with potential/kinetic ratio $r_s$ and density $\rho$ from
    $r_s = \frac{a}{a_0},
    a = \frac{1}{\sqrt{\pi \rho}},
    a_0 = \frac{\hbar^2}{m e^2}, 
    \textrm{Ry} = \frac{m e^4}{2\hbar^2}$.
The lower single-particle band of $\lambda$-jellium has quadratic dispersion $\epsilon_{\v{q}} = q^2/r_s^2$ where $q = |\v{q}|$ as usual. The wavefunction of the bottom band is $    \phi_{\v{q}} = \frac{1}{\sqrt{1 + \lambda^2 |\v{q}|^2}}
    \begin{pmatrix}
        1 \\
        \lambda(q_x + iq_y)
    \end{pmatrix},$
which has a skyrmionic texture in momentum space: the spinor points up at $\v{q}=0$, and winds the Bloch sphere once and points down as $\v{q}\to \infty$.

The texture carries Berry curvature $\Omega(\v{q}) = 2 \lambda^2 [\lambda^2 \n{q}^2 + 1]^{-2}$ where as usual the second band is pushed up above energy $\Delta \to \infty$, and is irrelevant. This gives a quadratic band independently adjustable potential/kinetic ratio $r_s$ and Berry curvature concentration $\lambda$, as claimed. Fig.~\ref{fig:lambda_jellium_phase_diagram} shows the mean-field phase diagram of the $\lambda$-jellium away from small $r_s$. The anomalous Hall crystal phase occupies a large region of the phase diagram, where it competes with Wigner crystals at large $r_s$. For a detailed tour of the mean-field phase diagram, we refer the readers to Ref.~\cite{soejima2025jelliummodelanomaloushall}.

\begin{figure}[h]
    \centering
    \includegraphics[width=\linewidth]{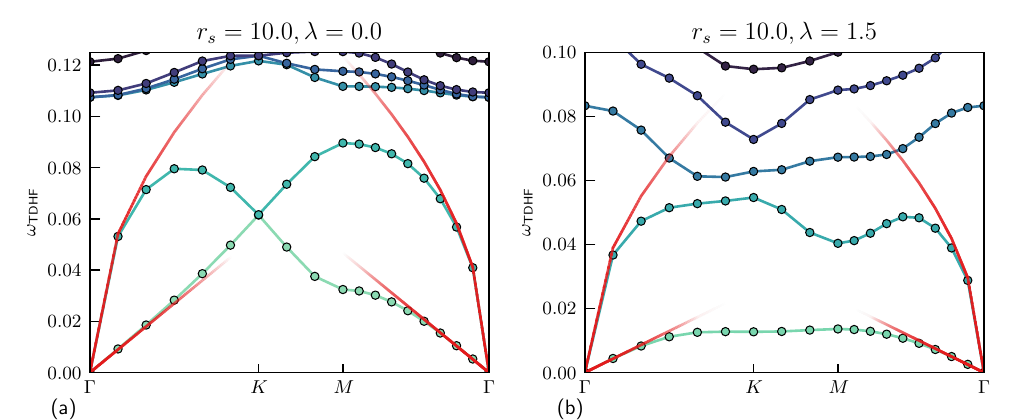}
    \caption{Plasmon spectrum of (a) the Wigner crystal and (b) the anomalous Hall crystal in the $\lambda$-jellium model. Dots are time-dependent Hartree-Fock spectra computed on a $18\times 18$ system with $7$ bands taken. Red lines come from the effective phonon action with coefficients from linear response. The two methods display excellent quantitative agreement.}
    \label{fig:WC_AHC_TDHF}
\end{figure}
\subsubsection{Benchmarking Assessment: Case of Wigner Crystals}
As a sanity check on our linear response methods, we focus first on $\lambda=0$. There   Eq.~\ref{eq:lambda_Jellium_Hamiltonian} reduces to the standard jellium model. Above $r_s =3$ the mean field ground state is a triangular lattice Wigner crystal whose elastic theory is well-understood~\cite{Bonsall-Maradudin}. Galilean invariance of the Hamiltonian fixes the effective mass to be $m = r_s^2/2$, while the shear stiffness can be computed in classical electrostatics as $\mu \approx 0.276/{r_s} + O(r_s^{-3/2})$. The kineo-elastic term vanishes by time-reversal symmetry. 
Our linear-response elastic theory accurately describes the low-energy excitations of the Wigner crystal. Wigner crystals have two gapless collective modes: the transverse and longitudinal phonons whose dispersions are
\begin{equation}
        \omega_t(\v q) = \sqrt{\frac{\mu}{m}}|\v q|+\dots, \quad
        \omega_l(\v q) = \sqrt{\frac{V(\v q) |\v q|^2\rho}{m}}+\dots,
\end{equation}
respectively. For long-ranged Coulomb interactions $V(\v q) = 4\pi/r_s|\v q|$, leading to a standard plasmon dispersion $\omega_{\ell} \propto \sqrt{q}$, just as in Fermi liquids. Fig.~\ref{fig:WC_AHC_TDHF}(a) shows the dispersions $\omega_{t,l}$ predicted by the phonon effective action using the linear response coefficients, as well as the neutral excitation spectrum computed using time-dependent Hartree-Fock (TDHF) computations. Their agreement confirms the validity of our procedure.

\begin{figure*}[htbp]
    \centering
    \includegraphics[width=\linewidth]{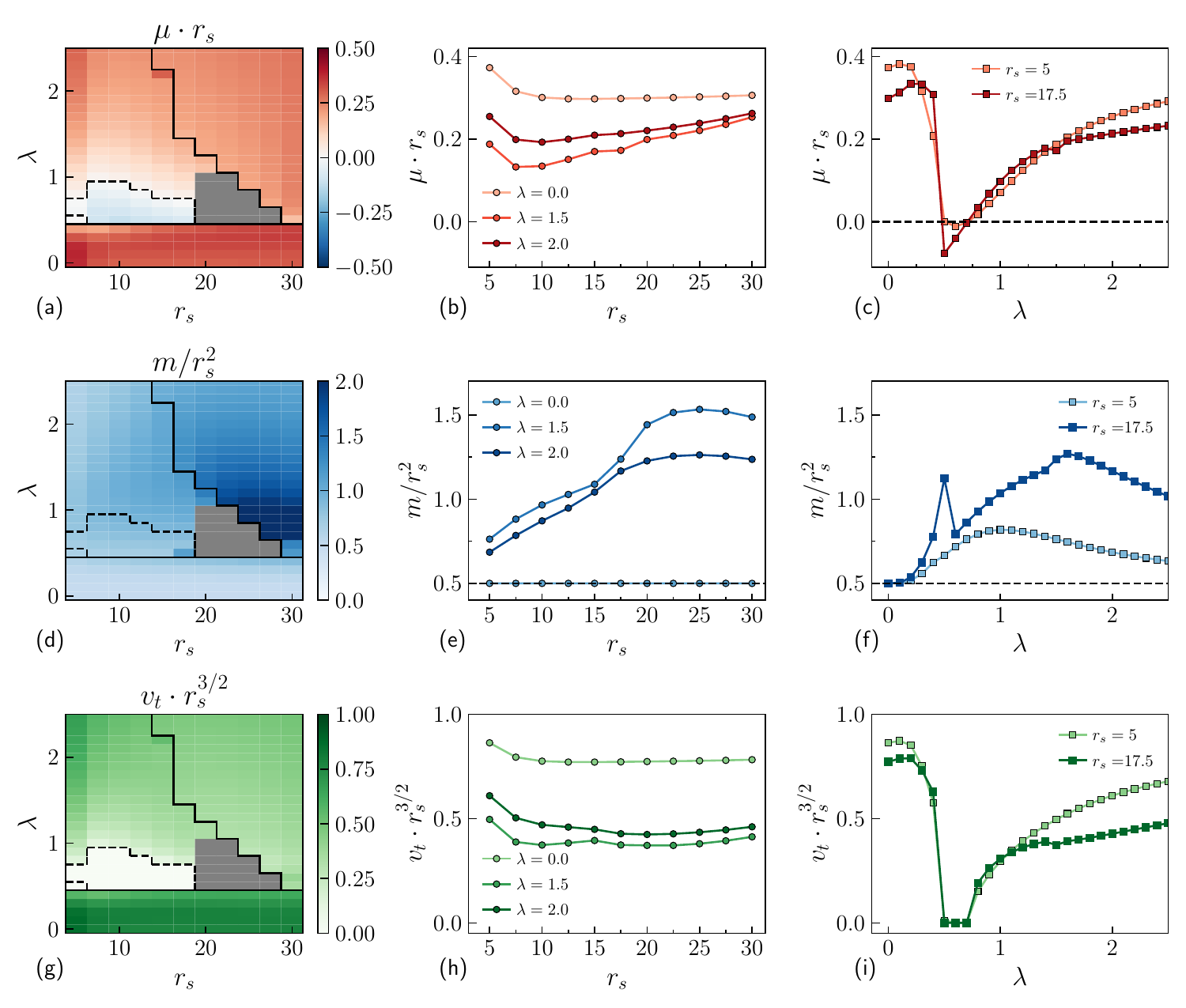}
    \caption{
    Numerical elastic parameters for the low-energy phonon theory of $\lambda$-jellium. 
    Panels (a-c) show shear stiffness, panels (d-f) show effective mass, and panels (g-i) show velocity. Quantities are scaled by appropriate powers of $r_s$ to make the large $r_s$ limit order unity (see text). 
    (a) Scaled shear stiffness $\mu \cdot r_s$ as a function of $\lambda$ and $r_s$. The gray region suffers from convergence issues,
    precluding accurate determination of the stiffness.
    Solid lines mark the phase boundary of different crystals, and the dotted lines surround a region of negative stiffness.
    (b) Line cuts of scaled shear stiffness against $r_s$. 
    (c) Line cuts of scaled shear stiffness against $\lambda$. Shear stiffness decreases precipitously at the first-order transition between WC and AHC, becoming negative in a small region (see text).
    (d) The scaled effective mass $m/r_s^2$ plotted against $r_s$ and $\lambda$. 
    (e) Line cut of the scaled effective mass versus $r_s$, showing interaction effects enhances the mass at nonzero $\lambda$.
    (f) Line cut of the normalized effective mass versus $\lambda$. An intermediate value of $\lambda$ enhances effective mass drastically.
    (g) Scaled speed of sound $v_t \cdot r_s^{3/2}$ plotted against $r_s$ and $\lambda$. 
    (h) Line cut of the scaled speed of sound against $r_s$. 
    (i) Line cut of the scaled speed of sound against $\lambda$. All data are computed from SCHF ($18\times18$) with $13$ bands taken, where a triangular unit cell was assumed.
    }
    \label{fig:lambda_jellium_results}
\end{figure*}

\subsubsection{Soft and Heavy Crystals with Berry Curvature}  

We now add Berry curvature to the mix, studying $\lambda$-Jellium at $\lambda > 0$. There are two other crystalline states in that regime: a large anomalous Hall crystal phase that eventually undergoes a \textit{continuous} transition into a second Wigner crystal phase at large $(r_s,\lambda)$. In fact this second Wigner crystal has different angular momentum than the usual Wigner crystal, and its momentum occupations have a ``halo" pattern around $\v{q}=0$; we call it the ``halo" Wigner crystal as was discussed in detail in Ref.~\cite{soejima2025jelliummodelanomaloushall} (see also Ref.~\cite{joy2023wignercrystallizationbernalbilayer}).

When $\lambda > 0$, the Hamiltonian lacks Gailean invariance, allowing the mass term to vary. However, inversion symmetry remains, forcing the kineo-elastic term to vanish. We now examine the mass and stiffness of the crystalline states in this regime. We begin by studying triangular unit cells, but will soon find we must broaden our choice of lattice.

Fig.~\ref{fig:lambda_jellium_results} shows the linear-response stiffness $\mu$, mass $m$ and the derived speed of sound $\sqrt{\mu/m}$ over the $\lambda$-jellium phase diagram. For each we show the phase diagram and line cuts of interest, which we now step through in sequence.

Stiffness is shown in Fig.~\ref{fig:lambda_jellium_results}(a-c). At asymptotically large $r_s$, the stiffness generally scales as $1/r_s$; we therefore plot $\mu \cdot r_s$. Fig.~\ref{fig:lambda_jellium_results}(b) shows linecuts at constant $\lambda$, which show the crystal softens considerably at intermediate $\lambda$ before recovering at large $\lambda$ to values close to those of the Wigner crystal.

In fact, the stiffness changes drastically upon the first order phase transition~\cite{soejima2025jelliummodelanomaloushall} from the WC1 to AHC, and becomes negative (Fig.~\ref{fig:lambda_jellium_results}(c)).
The negative shear stiffness(the dashed region in Fig.~\ref{fig:lambda_jellium_results}(a)) signals an instability of the triangular AHC, which we analyze in detail in Sec.~\ref{sec:two_band_energy_landscape}.
The triangular AHC phase becomes stable above $\lambda=0.75-1$, depending on $r_s$, where the stiffness becomes positive. The stiffness increases monotonically with $\lambda$ beyond that point, as it undergoes a continuous phase transition to halo WC, eventually recovering close to the WC value.

This can be understood from the single-particle wavefunctions of the $\lambda$-jellium. The electrons have a concentrated Berry curvature in a region of size $1/\lambda$ around the origin.  On the other hand, most of the electrons outside the region have spin down, and therefore trivial. When $\lambda$ becomes large, the trivial electrons dominate the energetics, and they crystalize into a triangular crystal as if they are in the jellium model. This claim is corroborated by the observation that the shear stiffness of the triangular crystal is smallest in the AHC phase when $\lambda$ is small.

In Fig.~\ref{fig:lambda_jellium_results}(d-f), we show the effective mass.
Nonzero $\lambda$, which breaks Galilean invariance of the projected interaction, enhances the effective mass relative to the Galilean invariant value $r_s^2/2$.
Intriguingly, the mass increases upon entering the halo WC, reaching up to four times the Galilean value.

The small stiffness and large mass at large $\lambda$ conspire to make the phonon modes slow. We show the velocity of the transverse phonon mode, estimated by $\sqrt{\mu/m}$ in Fig.~\ref{fig:lambda_jellium_results}(g-i).
This estimate for the transverse velocity, as well as that of the plasmonic dispersion matches TDHF calculations well in the small $\v{q}$ limit (Fig.~\ref{fig:WC_AHC_TDHF}(b)). Beyond acoustic phonons, the optical phonons of the AHC are significantly more dispersive than their WC counterparts. The origin of such behavior is worthy of future investigation.

\begin{figure*}[htbp]
    \centering
    \includegraphics[width=\linewidth]{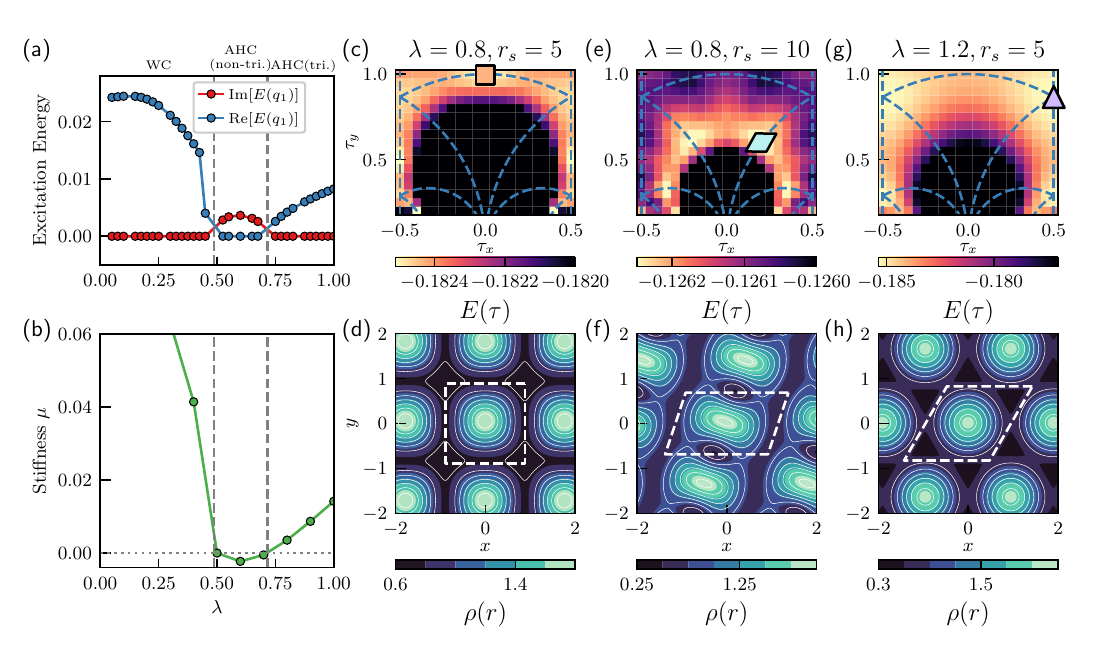}
    \caption{
    The shape of anomalous Hall crystal (AHC) depends on microscopic parameters. (a,b): for $r_s=5$, the triangular lattice is locally unstable when $\lambda\in[0.5,0.7]$. Non-zero imaginary part in the TDHF collective mode spectrum at $\v q_1 = \v G_1/12$ and the negative shear modulus both point to instability.
    (c): Energy landscape for different lattices parameterized by $\tau$ at $\lambda=0.8, r_s=5$. The energetically preferred lattice is the square lattice where $\tau=i$. (e): Energy landscape for $\lambda=0.8, r_s=10$. The energetically preferred lattice is a rhombic lattice where $\tau=0.2+0.6i$. (f): Energy landscape for $\lambda=1.2, r_s=5$. The energetically preferred lattice is the triangular lattice.
    All data are computed from SCHF ($24\times24$) with $13$ bands taken.
    }
    \label{fig:landscape}
\end{figure*}

\subsubsection{AHC Energy Landscape for Different Lattice Choices}
\label{sec:two_band_energy_landscape}

We now undertake a comprehensive study of the stability and instability of the triangular unit cell geometry in $\lambda$-jellium. As mentioned above, the stiffness of the triangular lattice AHC state is negative in a region near the WC phase boundary, enclosed by dashed lines in Fig.~\ref{fig:lambda_jellium_results}(a). A negative stiffness measurement implies the self-consistent Hartree-Fock ground state --- which is \textit{constrained} to use a triangular unit cell --- is not even a local minimum in the energy landscape. As a crosscheck, we perform TDHF calculations at $r_s = 5$ as a function of $\lambda$ (Fig.~\ref{fig:landscape}(a,b)). In the same regime where the stability is negative, we find the TDHF spectrum becomes imaginary, which occurs when the Hessian of the Hartree-Fock energy functional over the space of Slater determinants is not positive definite, which again implies the state is not at a local minimum.

This is consistent with our earlier Hartree-Fock work, which found that the square lattice AHC is energetically favored over the triangular lattice AHC in this same region, in stark contrast to the classical Wigner crystal where the square is disfavored~\cite{soejima2025jelliummodelanomaloushall}. This leads to an obvious question: is the AHC phase itself destabilized, or does it merely choose a different unit cell? To settle this, we now perform a comprehensive analysis of the landscape of unit cell shapes to find its \textit{global} minimum.

As our model is rotationally symmetric, the space of unit cell shapes parametrized by the modular parameter $\tau$ of the torus, whose fundamental domain is shown in Fig.~\ref{fig:methodology}(b). The size is fixed by requiring one electron per unit cell. Fig.~\ref{fig:landscape}(c-h) shows the $\tau$ energy landscape. At $\lambda=1.2$, Fig.~\ref{fig:methodology}(g,h), the minimum is at $\tau = e^{i2\pi/3}$, showing the triangular lattice AHC is not just a local minimum but the global minimum there. (One can see the landscape is indeed modular invariant.) At $\lambda = 0.8$ the landscape becomes extremely flat: the relative energy difference between the square lattice and the triangular lattice is around $0.02\%$ for both $r_s=5$ and $r_s=10$. By comparison, in the classical limit the triangular Wigner crystal is preferred over the square WC by $0.5\%$ --- an order of magnitude larger.
It is therefore no surprise that the global minimum shifts significantly with $r_s$, and is likely sensitive to other details. For $r_s =5$, the square lattice is indeed the global minimum, Fig.~\ref{fig:landscape}(c,d). Meanwhile at $r_s= 10$, the energetically preferred lattice is rhombic, Fig.~\ref{fig:landscape}(e,f), with $\tau=0.2+0.6i$ (note $|1-\tau|=1$.) As $\lambda$ increases, the triangular lattice is increasingly favored, becoming the global minimum by $\lambda\approx 1$. The precise shape of the minimum energy unit cell depends sensitively on physical details; the robust feature of these landscape is their flatness.

\subsection{Rhombohedral Multilayer Graphene}
\label{sec:RMG}
In this section, we turn to rhombohedral multilayer graphene (RMG), an experimental platform for which the anomalous Hall crystal was initially proposed by us and collaborators~\cite{AHC1}, along with simultaneous works~\cite{AHC_Yahui, AHC_Senthil}. Recent works have drawn attention to the different possible choices of lattice for AHC phases in RMG, in particular the possible instability of a triangular AHC with one electron per unit cell~\cite{MoireFCI3,zhou2024newclassesquantumanomalous,YBK_elastic}. Here, we point out that the stability of such an AHC is dependent on microscopic parameters. For instance, the orientation of the emergent lattice is fixed by the trigonal warping of RMG,
which breaks continuous rotation symmetry down to $C_3$ rotation symmetry. Also, and more importantly, we highlight that when a valley polarized crystal forms -- i.e. when time reversal and inversion are both broken -- the kineo-elastic coupling $\ell_{abc}$ is generally present. This coupling leads to surprising consequences for the collective modes of the system, such as a giant anisotropy in the phonon speeds along principle axes.

We consider a standard microscopic model of rhombohedral pentalayer graphene with screened Coulomb interactions, detailed in 
App. 1. For definiteness, we focus on a low filling $n\approx 1.06\times 10^{12}\si{cm^{-2}}$ and high displacement field $u_D=40\si{meV}$, whose flat valence band minimum contains significant Berry curvature. For comparison, this density would fill one band of a graphene-hBN moir\'e system with twist angle $0.9^\circ$. On the triangular lattice, the mean field ground state (polarized to a single spin and valley) is an anomalous Hall crystal. Unlike in $\lambda$-jellium, its energy depends on the relative orientation between the electronic crystal and the carbon lattice; we find the lowest energy configuration has relative orientation of $\phi = \pi/6$ between its crystal axis and $\v{a}_1$ of graphene.
\footnote{We note that Ref.~\cite{desrochers2025elasticresponseinstabilitiesanomalous} has found that the triangular AHC is unstable at a slightly smaller electronic density with $u_D=36\si{meV}$; this is likely due to a difference in microscopic modeling. Furthermore, the same reference finds that when the triangular lattice is aligned with the graphene lattice, the energy is minimized. We find that in contrast that with our parameters it is most stable when there is a $30$-degree relative rotation between them.}
Within TDHF, detailed below, this $\pi/6$ anomalous Hall crystal has a purely real collective mode spectrum across the entire Brillouin zone. We conclude that this mean-field triangular anomalous Hall crystal is locally stable.

\begin{figure*}[htbp]
    \centering
    \includegraphics[width=0.95\textwidth]{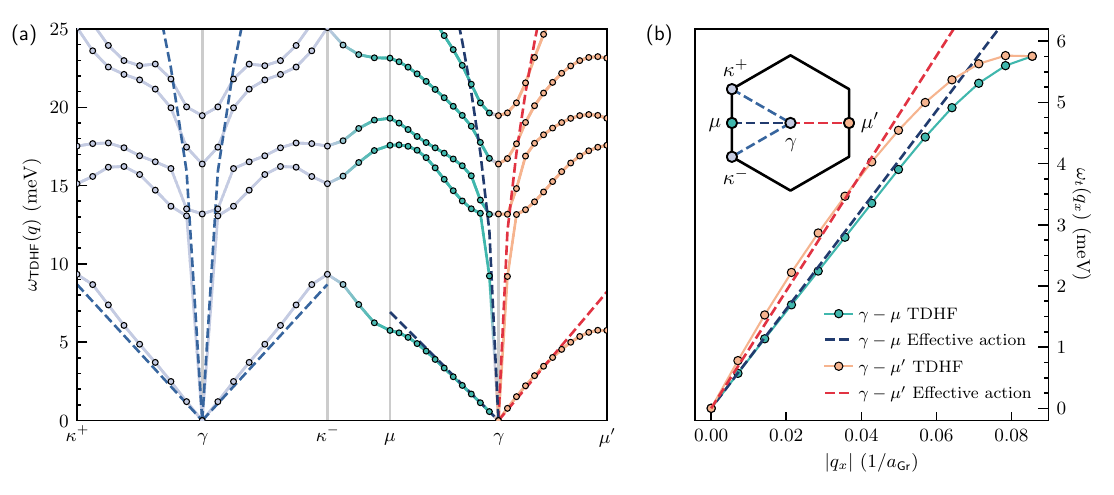}
    \caption{Time-dependent Hartree-Fock results for a triangular anomalous Hall crystal in rhombohedral multilayer graphene. Parameters: electronic density $n=1.06\times 10^{12}\si{cm^{-2}}$, displacement field $u_D=40\si{meV}$, relative orientation between the electron lattice and graphene lattice $\phi=\pi/6$, and relative dielectric constant $\epsilon_r=5$. (a) TDHF spectrum $\omega(\v q)$ along a high-symmetry line cut (dots), computed using a $24\times 24$ torus. Colors represent values of $\cos(3\theta_{\v q})$ where $\theta_{\v q}$ is the angle between $\v q$ and the $x$ axis. Dashed lines correspond to phonon dispersions predicted by effective field theory with parameters extracted numerically: $\mu\approx 1.8\si{meV},m\approx 0.29m_e,\ell\approx 1.3\times 10^{-2}\si{nm}^{-1}$.
    (b) Due to nonzero kineo-elastic coupling $\ell$, the phonon speeds differ by $6\si{km/s}$ along the $x$ axis ($\mu-\gamma-\mu'$ line). All data are computed from SCHF ($24\times24$) with $7$ bands taken. For detailed estimates of physical quantities refer to Tab.~\ref{tb:RMG_stable_point}.} 
    \label{fig:RMG_TDHF}
\end{figure*}

Given this local stability, we proceed to examine the triangular AHC phonon spectrum. Fig.~\ref{fig:RMG_TDHF}(a) compares the phonon spectrum of the effective field theory, with linear response elastic coefficients from   Eq.~\ref{eq:c3_extraction}, to time-dependent Hartree-Fock numerics. As expected, we find quantitative agreement between the techniques at low energies for both transverse and longitudinal phonons.

\begin{table}[h]
 \renewcommand{\arraystretch}{1.3}
 \centering
\begin{tabular}{ lc } 
 \toprule
  Physical Quantity & Numerical Value  \\ \hline \\[-1.2em]
  Electronic density $\rho$ & $1.06\times 10^{12}\si{cm^{-2}}$  \\ 
  Displacement field $u_D$ & $40 \si{meV}$ \\ 
  Dielectric constant $\epsilon_r$ & $5$ \\
  Relative orientation & $30^\circ$ \\
  Chern Number $\n{C}$ & $1$ \\
  Charge gap & $15 \si{meV}$ \\
  Shear stiffness $\mu$ & $1.8\si{meV}$ \\
  Effective mass $m$ & $0.29m_e$ \\
  Kineo-elastic coupling $\ell/\hbar$ & $1.3\times 10^{-2}\si{nm^{-1}}$\\
  Transverse speed $v_{+x}$ & $37 \si{km/s}$ \\
  Transverse speed $v_{-x}$ & $31 \si{km/s}$\\
  Transverse speed $v_y$ & $33\si{km/s}$\\
  Melting temperature $T_c$ & $1.9\si{\kelvin}$\\
 \bottomrule
\end{tabular}
\caption{Physical properties of the anomalous Hall crystal in a microscopic model of rhombohedral pentalayer graphene.}
\label{tb:RMG_stable_point}
\end{table}

Table~\ref{tb:RMG_stable_point} shows the elastic parameters and other physical properties of the stable triangular anomalous Hall crystal --- including a non-zero kineo-elastic coupling. Since the AHC is valley polarized, it breaks both inversion and time reversal symmetry. Thus the kineo-elastic terms $\ell_{abc}$ are symmetry allowed. As discussed in Sec.~\ref{sec:elastic_theory}, the kineo-elastic tensor has only two independent components in the presence of $C_3$ symmetry:
\begin{equation}
\begin{aligned}
    \ell_{abc}u_a\partial_b u_c &= \ell(\dot{u}_{x}(u_{xx}-u_{yy})-\dot{u}_y(2u_{xy})) \\&+ \ell'(\dot{u}_{y}(u_{xx}-u_{yy})+\dot{u}_x(2u_{xy})).
\end{aligned}
\end{equation}
Furthermore, $M_x \mathcal{T}$ symmetry acts within a valley, implying $\ell'=0$. To measure the single remaining component $\ell$, we apply shear strain $\theta_{xy}$, which we find induces a non-zero momentum in the $y$-direction (fixed by $M_x \mathcal{T}$). Computing the resultant $\braket{\hat{p}_y}$ easily yields $\ell$ via   Eq.~\ref{eq:c3_extraction}.

The presence of kineo-elastic coupling $\ell$ significantly alters the phonon dispersion. 
It produces a large anisotropy in the transverse phonon speeds
\begin{equation}
    v_{\pm} = \frac{\sqrt{4m\mu+\ell^2}\pm \ell}{2m}
\end{equation}
along the $\pm x$ directions, as discussed around Eq.~\ref{eq:phonon_spectrum_and_velocity}. Fig.~\ref{fig:RMG_TDHF}(b) shows this clear asymmetry along the the $\mu-\gamma-\mu'$ line: $v_{+}-v_{-} = 6\si{km/s}$, a $20\%$ relative difference. Conversely, there is no asymmetry along the $y$ direction and its $C_3$ images due to $M_x \mathcal{T}$; the transverse velocity along the $\kappa^+ -\gamma -\kappa^-$ line is $v_t=\sqrt{\mu/m}\approx 33\si{km/s}$.

The longitudinal phonons are unaltered by the kineo-elastic coupling. Our Coulomb interactions are screened beyond the gate distance $d$, so the longitudinal phonons are linear up to $q \approx 1/d$, whereupon they crossover to the characteristic plasmon-like behavior. We take $d = \SI{250}{\r{A}}$, on the order of the smallest momentum transfer in our $24 \times 24$ system, which produces an approximately $\sqrt{q}$ dispersion throughout the Brillouin zone. Furthermore, we observe a series of avoided crossings between the longitudinal phonons and higher bands of collective modes, which are beyond the low-energy effective theory.

We can further estimate the melting temperature of the crystal, assuming that it is governed by defect proliferation as described by the KTHNY theory ~\cite{KTMelting_A,KTMelting_B,NelsonKosterlitz,NelsonMelting,FisherShear}. The transition temperature in this case is governed by the temperature renormalized stiffness $\mu(T)$. Approximating it by the zero temperature stiffness, we find
\begin{equation}
    T_c=\frac{\mu(T=T_c)}{2\pi\sqrt{3}}\simeq\frac{\mu(T=0)}{2\pi\sqrt{3}}=0.17\si{\milli\electronvolt}=1.9\si{\kelvin},
\end{equation}
which means the crystal could be stable at experimentally relevant temperatures.

\section{Discussion}
\label{sec:discussion}
In this work, we derived on the lowest-order gradient expansion of the phonon effective action. Higher-order gradient terms can reveal other nontrivial features of the crystal. For example, Ref.~\cite{Zeng_sliding} found that the presence of anomalous velocity modifies the frequency of the phonon modes. We expect other quantum geometric effects to affect higher order terms in the phonon action as well, which we leave for future work.

Our parameter extraction scheme can be generalized beyond Hartree-Fock methods.
Two directions are particularly natural: (i) The Hartree-Fock approximation can be viewed as the lowest order conserving approximation to the true effective action, within the 2PI formalism~\cite{Baym2PI} (See also App.~3). This approximation can be improved by keeping higher order diagrams in the Luttinger-Ward functional.
(ii) We can use improved numerical methods for ground state energy estimation, such as variational Monte-Carlo methods.
Our approach gives us the ability to access dynamical information from ground state energies alone. 

The kineo-elastic term $\ell_{abc}$ we proposed renders the phonons non-reciprocal~\cite{Cheong2018BrokenSymmetries,Tokura2018Nonreciprocal}.
Previous studies of non-reciprocal phonons have focused on $\mathcal{PT}$-symmetric antiferromagnets~\cite{ren2024nonreciprocalphononsptsymmetricantiferromagnet}, the phonon magneto-chiral effect where a magnetic field is applied to a chiral material~\cite{Nomura2019Phonon,Aztori2021Magneto,Nomura2023Nonreciprocal}, metamaterials~\cite{TopologicalPhononics1,TopologicalPhononics2,Nassar2020Metamaterial}, or surface acoustic waves~\cite{Xu2020Nonreciprocal,Shao2021Electrical}. In particular, to the best of our knowledge, such a kineo-elastic term was not previously proposed in the literature, although magnon-phonon coupling could induce such an effect~\cite{DiXiao_phonons}~\footnote{We thank Di Xiao for pointing out to us that in such scenarios the kineo-elastic coupling is several orders of magnitudes smaller due to the magnon-phonon coupling and thus usually ignored.}; prior discussions of non-reciprocal effects have focused on higher gradient terms~\cite{arora2022quantum,dutta2023intrinsic,ren2024nonreciprocalphononsptsymmetricantiferromagnet}, which do not affect the speed of sound at $\v q=0$. In contrast, the kineo-elastic term dramatically modifies the speed of sound.

Finally, we comment on the implications of this work for rhombohedral graphene systems. The kineo-elastic term will exist in the phonon effective action whenever the electronic crystal is valley polarized.  Thus, the anisotropic speed of sound can probe ``valley polarization'', even within the high-resisitivity phase which is observed at low densities of rhombohedral graphene~\cite{lu2024fractional}, a putative  Wigner crystal,  whose valley ordering remains an open question. We also find that the triangular anomalous Hall crystal can become unstable for other microscopic parameter choices, and indeed Refs.~\cite{kwan2023moir,YBK_elastic} have reported the triangular lattice is destabilized under other parameters. One possibility is that, just as in $\lambda$-jellium, the global mean-field ground state at those parameters is also an anomalous Hall crystal --- but with a potentially different orientation and shape, such as an oblong rectangle. Exploring the unrestricted landscape of unit cell shapes, orientations, and even expanding the unit cell area is an important direction for future work.

\begin{acknowledgements}
We thank Taige Wang, Tianle Wang, Mike Zaletel, Patrick J. Ledwith, and Eslam Khalaf for related collaborations and useful insights. 
We also thank Erez Berg, Yaar Vituri, Agnes Valenti, Miguel Morales,  Shiwei Zhang, Ethan Lake, Dam Thanh Son, Haruki Watanabe, Leon Balents, Bert Halperin, Tarun Grover, Yafei Ren, Chong Wang, Di Xiao, Xiao-Wei Zhang, Yongxin Zeng, F\'{e}lix Desrochers, and Yong Baek Kim for fruitful discussions.
This research is funded in part by the
Gordon and Betty Moore Foundation’s EPiQS Initiative,
Grant GBMF8683 to T.S.; A.V., O.E.S. and J.D. were funded by NSF DMR-2220703.
AV is supported by the Simons Collaboration on Ultra-Quantum Matter, which is a grant from the Simons Foundation (651440, A.V.).
D.E.P. acknowledges startup funds from UC San Diego.
\end{acknowledgements}

\bibliographystyle{unsrt}
\bibliography{references}

\onecolumngrid
\appendix
\newpage

\section{Microscopic Hamiltonian of rhombohedral multilayer graphene}
\label{app:RMG}

This Appendix reviews the microscopic Hamiltonian of rhombohedral multilayer graphene. We follow the conventions from Ref.~\cite{AHC1}. Here, we provide the minimal summary necessary for reproducing the Hamiltonian.

The direct lattice for graphene is given by 
\begin{align}
	\v{R}_1 = a\left(1,0 \right), \qquad
	&\v{R}_2 = a\left( \frac{1}{2}, \frac{\sqrt{3}}{2} \right),
\end{align}
where we take the lattice constant to be $a \approx 0.246\si{nm}$. 

The Hamiltonian for rhombohedral multilayer graphene~\cite{zhang2010band,jung2013gapped} (see also:~\cite{chen2019signatures,zhou2021half,chatterjee2022inter,ghazaryan2023multilayer,park2023topological} and references therein) in Fourier space is a $2\ell \times 2\ell$ dimensional matrix with the block structure
\begin{equation}
	h_{RG}^{(N_L)}(\v{k}) =
	\begin{bmatrix} 
		h^{(0)}_{\ell} & h^{(1)} & h^{(2)}\\[0.5em]
		h^{(1)\dagger} & h^{(0)}_{\ell} & h^{(1)} & h^{(2)}\\[0.5em]
		h^{(2)\dagger} & h^{(1)\dagger} & h^{(0)}_{\ell} & \ddots & \ddots\\[0.5em]
				& h^{(2)\dagger} & \ddots & \ddots & h^{(1)} & h^{(2)}\\[0.5em]
				& & \ddots & h^{(1)^\dagger} & h^{(0)}_{\ell} & h^{(1)}\\[0.5em]
				& & & h^{(2)\dagger} & h^{(1)\dagger} & h^{(0)}_{\ell}\\
	\end{bmatrix}
	\label{eq:rhombohedral_graphene}
\end{equation}
whose subblocks are parameterized as
\begin{subequations}
	\begin{align}
		h^{(0)}_{\ell} (\v{k}) &= \begin{pmatrix} u_{A\ell}  & -t_0 f_{\v{k}}\\ -t_0 \overline{f}_{\v{k}} & u_{B\ell} \end{pmatrix},
		&h^{(1)}(\v{k}) = \begin{pmatrix} t_4 f_{\v{k}} & t_3 \overline{f_{\v{k}}} \\ t_1 & t_4 f_{\v{k}} \end{pmatrix},\qquad
		&h^{(2)}(\v{k}) = \begin{pmatrix} 0 & \frac{t_2}{2}\\ 0 & 0 \end{pmatrix}, 
		&f_{\v{k}} = \sum_{i=0}^2 e^{i \v{k}\cdot \v{\delta}_i},
	\end{align}
\end{subequations}
where we defined the nearest neighbor vectors to be $\v{\delta}_{n} = R_{n 2\pi/3} (0,\frac{1}{\sqrt{3}}a)^T$ for $n=0,1,2$ where $R_{\theta}$ is the counterclockwise rotation matrix by angle $\theta$.

For microscopic parameters, we take parameters extracted from Ref.~\cite{wang2023electrical}:
\begin{equation}
    \begin{aligned}
        t_0 &= 3100\si{meV},\\
        t_1 &= 380\si{meV},\\
        t_2 &= -21\si{meV},\\
        t_3 &= 290\si{meV},\\
        t_4 &= 141\si{meV}.
    \end{aligned}
\end{equation}
The displacement field is implemented as a uniform gradient of potential energy throughout the graphene layers:
\begin{equation}
	u_{\sigma, \ell} = u_D \left( \ell + 1 - \frac{N_L-1}{2} \right).
\end{equation}

The full many-body Hamiltonian is given by
\begin{equation}
    \hat{H} = \hat{h}_{RG}^{(N_L=5)} + \hat{V}
\end{equation}
where $\hat{V}$ encodes standard double-gate screened Coulomb interactions
\begin{equation}
    V(q) = \frac{2\pi \tanh q d}{\epsilon_r \epsilon_0 q},
\end{equation}
where $\epsilon_r =5$ is the relative dielectric constant and $d = \SI{250}{\r{A}}$ is the gate distance. We use the ``charge neutrality" subtraction scheme~\cite{AHC1}.

Our self-consistent Hartree-Fock calculations are projected into seven bands with system sizes up to $24 \times 24$. We consider a single valley and spin only. When performing time-dependent Hartree-Fock, the mean-field correlation matrix $P$ was converged so that the commutator error obeys $[H_{HF}[P], P] = 0$ with accuracy close to machine precision. This avoids spurious imaginary eigenvalues in the collective mode spectrum. The TDHF eigenvalue problem is solved using standard Arnoldi methods \textit{without} shift-inversion. We ensure good convergence in the case of degenerate eigenvalues, such as the exact zero modes at $\gamma$.

\section{TDHF}

This Appendix derives the time-dependent Hartree-Fock formulation of particle-hole excitations~\cite{blaizot_quantum_1986}. We follow the notation and presentation of Appendix A of~\cite{khalaf2020soft}, with some use of~\cite{nick_notes}. The main goal is to derive formulas that are well-suited to numerical calculations using band-projected Hartree-Fock.

\subsection{The Idea of TDHF}

Time-dependent Hartree-Fock is a special case of the time-dependent variational principle, a lens through which the procedure becomes simple. Given a Hamiltonian, we can compute the energy $E[P]$ of a Slater determinant $P$. To second order, the energy of nearby Slater determinants is (using a schematic notation) 
\begin{equation}
    E[P+ dP] = E[P] + J_{P} \cdot dP + \frac{1}{2} dP^\dagger \cdot \mathcal{H}_{P}  \cdot  dP,
\end{equation}
where $J$ is a Jacobian and $\mathcal{H}$ is a Hessian. Self-consistent Hartree-Fock minimizes the energy over the variational manifold of Slater determinants, producing a local minimum $E[P_0] = E_0$. As a local minimum, the Jacobian vanishes: $J_{P_0} = 0$, while the Hessian $\mathcal{H}_{P_0}$ controls the potential well. The low-frequency dynamics of any small perturbation around the minima,
\begin{equation}
    P(t) = P_0 + \delta P(t),
\end{equation}
are therefore controlled by the smallest eigenvalues of the Hessian at the minima. Our goal is to compute them. Instead of using the positive semi-definite Hessian directly, it is more convenient in practice to use a slightly different operator with a particle-hole-like structure that is called $\mathcal{L}$ below. 

\subsection{Setup}

Consider a Hamiltonian
\begin{equation}
    \hat{H} = \hat{h} + \hat{V} = \sum_{k} \hat{c}^\dagger_{\v{k},a} \; [h(\v{k})]_{ab} \; \hat{c}_{\v{k},b}
    + \frac{1}{2A} \sum_{\v{q}}
    V_q \hat{\rho}_{\v{q}} \hat{\rho}_{-\v{q}}
\end{equation}
with $\hat{\rho}_{\v{q}} = \sum_{k} \hat{c}^\dagger_{\v{k},a} \; [\Lambda_{\v{q}}(\v{k})]_{ab} \; \hat{c}_{\v{k}+\v{q},b}$ as normal. Let
\begin{equation}
    \widehat{\phi}_{\v{q}} = \sum_{\v{k},a,b}
    \hat{c}^\dagger_{\v{k}} \; [\phi_{\v{q}}(\v{k})]_{ab} \; \hat{c}_{\v{k}+\v{q}}
\end{equation}
be a neutral collective mode where $\phi_q$ (no hat) denotes its corresponding matrix. Its true equation of motion is 
\begin{equation}
    i \partial_t \hat{\phi}_{\v{q}} = [\hat{H},\hat{\phi}_{\v{q}}] = \L \hat{\phi}_{\v{q}},
\end{equation}
where $\L = [\hat{H},\cdot]$ is the Liouvillian. We study the collective mode within TDHF over a particular mean field ground state with correlation matrix $P$, which is equivalent to the Bethe-Saltpeter collective mode at RPA level~\cite{nick_lecture}.

The TDHF collective modes are eigenvectors~\cite{khalaf2020soft}
\begin{equation}
    i \partial_t \hat{\phi}_q = \L^{\textsf{MF}}_{\v{q}} \hat{\phi}_{\v{q}} = \omega_{\v{q}} \hat{\phi}_{\v{q}}
\end{equation}
where $\L^{\textsf{MF}}_{\v{q}}$ is the ``mean-field Liouvillian" at momentum $\v{q}$. Since we wish to remain within the space of mean-field operators, we define
\begin{equation}
    \L^{\textsf{MF}}_{\v{q}} = \mathcal{P}^{\mathsf{MF}} \L,
\end{equation}
where $P_{\mathsf{MF}}$ projects an operator to mean-field level (with respect to a correlation matrix $P$) via
\begin{equation}
    \hat{O}^{\mathsf{MF}} = \mathcal{P}^{\mathsf{MF}}\hat{O} = \sum_{k} \hat{c}^\dagger_{\v{k},a} \; [O(\v{k},\v{k}')]_{ab} \; \hat{c}_{\v{k}',b}
    \end{equation}
    where
    \begin{equation}
    [O(\v{k},\v{k}')]_{ab} = \frac{\partial}{\partial P(\v{k},\v{k}')_{ab}} \braket{\hat{O}}_P.
    \label{eq:mean_field_projector}
\end{equation}
For instance,
\begin{equation}
    \hat{H}^{\mathsf{MF}} = \mathcal{P}^{\mathsf{MF}}\hat{H} = \hat{h}^{\mathsf{HF}}[P] = \hat{h} + \hat{h}_H[P] + \hat{h}_{F}[P],
\end{equation}
where these are these normal Hartree and Fock Hamiltonians.

We focus on particle-hole excitations only, i.e. those that that create or annihilates a PH excitation, i.e.
\begin{equation}
    \braket{\Psi_{\mathsf{MF}}|\hat{\phi}_{\v{q}}|\Psi_{\mathsf{MF}}} = 0.
\end{equation}
This is true whenever
\begin{equation}
    \hat{\phi}_{\v{q}} = \hat{P}^\perp \hat{\phi}_{\v{q}} \hat{P} \text{ or }
    \hat{\phi}_{\v{q}} = \hat{P}^\perp \hat{\phi}_{\v{q}} \hat{P} 
\end{equation}
where $\hat{P}$ is the projector onto the mean-field ground state and $\hat{P}^\perp = \hat{I} = \hat{P}$. Such states are in the image of the PH super-projector
\begin{equation}
    \mathcal{P}^{\mathsf{PH}}\hat{\phi}_{\v{q}}
    = \sum_{\v{k},a,b}
    \hat{c}^\dagger_{\v{k},a} \; [P^\perp(\v{k}) \phi_q(\v{k}) P(\v{k}+\v{q}) +  P(\v{k}) \phi_{\v{q}}(\v{k}) P^{\perp}(\v{k}+\v{q})]_{ab} \; \hat{c}_{\v{k}+\v{q},b}
\end{equation}
where $P^\perp = I - P$. These two terms create and destroy particle-hole excitations above the ground state, respectively.

We therefore wish to solve the (non-Hermitian) eigenvalue problem
\begin{equation}
    \L^{\textsf{PH}}_{\v{q}} \hat{\phi}_{\v{q}} = \omega_{\v{q}} \hat{\phi}_{\v{q}}
    \text{ where }
    \L^{\textsf{PH}} = \mathcal{P}^{\mathsf{PH}} \L^{\mathsf{MF}} \mathcal{P}^{\mathsf{PH}},
\end{equation}
which is an operator entirely within the space of mean-field particle-hole excitations over $P$.

\subsection{Boosted Hartree-Fock Hamiltonians}

The TDHF equation is best expressed in terms of two simple building blocks: the ``boosted commutator" and the ``boosted Hartree-Fock Hamiltonians".

Suppose $\hat{A} = \sum_{\v{k}} \hat{c}_{\v{k}}^\dagger A(\v{k}) \hat{c}_{\v{k}}$ and $\hat{B}_{\v{q}} = \sum_{\v{k}} \hat{c}_{\v{k}}^\dagger B_{\v{q}}(\v{k}) \hat{c}_{\v{k}+\v{q}}$. Here and below band indices will be implicit. Then
\begin{equation}
    [\hat{A},\hat{B}_{\v{q}}]
    = \sum_{\v{k}} \hat{c}_{\v{k}}^\dagger \; [A,B_{\v{q}}]_{\v{q}}(\v{k}) \; \hat{c}_{\v{k}+\v{q}}
\end{equation}
where the \textsf{boosted commutator} is
\begin{equation}
        [A,B_{\v{q}}]_{\v{q}}(\v{k})
    := A(\v{k}) B_{\v{q}}(\v{k}) - B_{\v{q}}(\v{k}) A(\v{k}+\v{q}).
\end{equation}

Now we consider the Hartree-Fock equations but with boosted correlation matrices, which have the form of $\hat{B}_{\v{p}}$. The \textsf{boosted Hartree Hamiltonian} is
\begin{align}
    \hat{h}_{H}[B_{\v{p}}]
    &=
2 \times \begin{tikzpicture}[scale=0.75,baseline={(0,1.5em)}]
\node[label=below:$\v{k}$] (i) at (-1.5,0) {};
\node[label=below:$\v{k}+\v{q}$] (f) at (1.5,0) {};
\node[dot,inner sep=1.5pt,label=below:$\Lambda_{\v{q}}$] (V1) at (0,0) {};
\node[dot,inner sep=1.5pt,label=above:$\Lambda_{-\v{q}}$] (V2) at (0,1.5) {};
\node[draw,rectangle,inner sep=2pt] (P) at (0,3) {$B_{\v{p}}$};
\draw[postaction={on each segment={mid arrow}}] (f) -- (V1) -- (i);
\draw[snake] (V1) -- (V2);
\draw[postaction={on each segment={mid arrow}}] (V2) -- ++(-0.5,0) to[bend left,out=90,in=90] (P.west)
(P.east) to[bend right,out=90,in=90]  ++(0.0,-1.5) -- (V2);
\node[rotate=90] at (-1.3,2.3) {$\v{k}'=\v{k}''+\v{p}$};
\node[rotate=270] at (1.3,2.3) {$\v{k}'-\v{q}=\v{k}''$};
\end{tikzpicture}
=\frac{1}{A} \sum_{\v{q}} \sum_{\v{k},\v{k}',\v{k}''} V_{\v{q}} \Lambda_{\v{q}}(\v{k})
       \tr[ \Lambda_{-\v{q}}(\v{k}') B_{\v{p}}(\v{k}'')] 
       \delta_{\v{k}',[\v{k}''+\v{p}]}
       \delta_{\v{k}'',[\v{k}'-\v{q}]}
       \hat{c}^\dagger_{\v{k}} c_{[\v{k}+\v{q}]}.
\end{align}
The total momentum around the loop must be zero $\pmod{\v{G}}$:
\begin{equation}
    \delta_{\v{k}',[\v{k}''+\v{p}]} \delta_{\v{k}'',[\v{k}'-\v{q}]}
    =  \delta_{\v{k}',[\v{k}''+\v{p}]} \sum_{\v{G}} \delta_{\v{q},\v{p}+\v{G}},
\end{equation}
so
\begin{align}
 \hat{h}_{H}[B_{\v{p}}]
=
\frac{1}{A} \sum_{\v{G}} \sum_{\v{k},\v{k}''} V_{\v{p}+\v{G}} \Lambda_{\v{p}+\v{G}}(\v{k}) \tr[ \Lambda_{-\v{p}-\v{G}}(\v{k}''+\v{p}) B_{\v{p}}(\v{k}'')]
\hat{c}^\dagger_{\v{k}} c_{[\v{k}+\v{p}+\v{G}]}.
\end{align}
Finally using $    \Lambda_{\v{q}}(\v{k}) = \Lambda_{-\v{q}}(\v{k}+\v{q})^\dagger$ we get
\begin{equation}
 \hat{h}_{H}[B_{\v{p}}]
=
\sum_{\v{k}} \hat{c}^\dagger_{\v{k}} h_H[B_{\v{p}}](\v{k}) c_{[\v{k}+\v{p}]}
\end{equation}
where
\begin{equation}
h_H[B_{\v{p}}](\v{k}) 
= \frac{1}{A} \sum_{\v{G}} V_{\v{p}+\v{G}} \Lambda_{\v{p}+\v{G}}(\v{k})
\sum_{\v{k}'} \tr[ \Lambda_{\v{p}+\v{G}}(\v{k}')^\dagger B_{\v{p}}(\v{k}')].
\end{equation}

Similarly, the \textsf{Boosted Fock Hamiltonian} is
\begin{align}
    \hat{h}_{F}[B_{\v{p}}] 
    =
\begin{tikzpicture}[scale=0.9,baseline={(0,1em)}]
\node (i) at (-2.75,0) {};
\node (f) at (1,0) {};
\node[draw,rectangle,inner sep=2pt] (P) at (-1,0) {$B_{\v{p}}$};
\node[dot,inner sep=1.5pt,label=below:$\Lambda_{\v{q}}$] (V1) at (-2,0) {};
\node[dot,inner sep=1.5pt,label=below:$\Lambda_{-\v{q}}$] (V2) at (0,0) {};
\draw[snake] (0,0) arc (0:180:1);
\draw[postaction={on each segment={mid arrow}}]
(f) -- (V2) -- (P.east)
(P.west) -- (V1) -- (i);
\end{tikzpicture}
= \frac{1}{A}\sum_{\v{q},\v{k},\v{k}',\v{k}''}
V_{\v{q}}
\Lambda_{\v{q}}(\v{k}) 
B_{\v{p}}(\v{k}')
\Lambda_{-\v{q}}(\v{k}'') \delta_{[\v{k}+\v{q}],\v{k}'}
\delta_{[\v{k'}+\v{p}],\v{k}''}
\hat{c}^\dagger_{\v{k}}
\hat{c}_{\v{k}''-\v{q}}.
\end{align}
We can eliminate $\v{k}' = [\v{k}+\v{q}]$ and $\v{k}'' =[\v{k}'+\v{p}] = [\v{k}+\v{q}+\v{p}]$ to get

\begin{equation}
        \hat{h}_{F}[B_{\v{p}}] =
 \frac{1}{A}\sum_{\v{q},\v{k},\v{k}',\v{k}''}
V_{\v{q}}
\Lambda_{\v{q}}(\v{k}) 
B_{\v{p}}([\v{k}+\v{q}])
\Lambda_{-\v{q}}(\v{k}+\v{p}+\v{q})
\hat{c}^\dagger_{\v{k}}
\hat{c}_{\v{k}+\v{p}+\v{q}-\v{q}}.       
\end{equation}
or
\begin{equation}
 \hat{h}_{F}[B_{\v{p}}]
=
\sum_{\v{k}} \hat{c}^\dagger_{\v{k}} h_F[B_{\v{p}}](\v{k}) c_{[\v{k}+\v{p}]}
\text{ where }
h_F[B_{\v{p}}](\v{k}) 
= \frac{1}{A} \sum_{\v{q}} V_{\v{q}}
\Lambda_{\v{q}}(\v{k}) 
B_{\v{p}}([\v{k}+\v{q}])
\Lambda_{\v{q}}(\v{k}+\v{p})^\dagger.
\end{equation}
In summary, we have found the \textsf{boosted Hartree-Fock Hamiltonian}
\begin{align}
    \hat{h}_{HF}[B_{\v{p}}]
    &= \sum_{\v{k}} \hat{c}^\dagger_{\v{k}} h_{HF}[B_{\v{p}}](\v{k})
    c_{[\v{k}+\v{p}]},\\
    h_{HF}[B_{\v{p}}]
    &=
    h_H[B_{\v{p}}] +    
    h_F[B_{\v{p}}],\\
h_H[B_{\v{p}}](\v{k}) 
&= \frac{1}{A} \sum_{\v{G}} V_{\v{p}+\v{G}} \Lambda_{\v{p}+\v{G}}(\v{k})
\sum_{\v{k}'} \tr[ \Lambda_{\v{p}+\v{G}}(\v{k}')^\dagger B_{\v{p}}(\v{k}')],\\
h_F[B_{\v{p}}](\v{k}) 
&= \frac{1}{A} \sum_{\v{q}} V_{\v{q}}
\Lambda_{\v{q}}(\v{k}) 
B_{\v{p}}([\v{k}+\v{q}])
\Lambda_{\v{q}}(\v{k}+\v{p})^\dagger.
\end{align}

\subsection{Derivation of TDHF}

We now reduce the TDHF operator $\mathcal{L}^{\mathsf{MF}}$ to an operator on matrices $\phi_{\v{q}}$. We therefore evaluate
\begin{equation}
    \L^{\textsf{MF}}_{\v{q}} \hat{\phi}_q
    = \mathcal{P}^{\mathsf{MF}}[\hat{h},\hat{\phi}_{\v{q}}] + \mathcal{P}^{\mathsf{MF}}[\hat{V},\hat{\phi}_{\v{q}}].
\end{equation}
We interpret Eq.~\eqref{eq:mean_field_projector} diagramatically: write all Wick contractions diagrams of $\hat{O}$ by $P$ with exactly two dangling legs (so that inserting one more $P$ would give $\braket{\hat{O}}_P$). The projector is trivial for the first term, yielding $[\hat{h},\hat{\phi}_{\v{q}}]$. For the second term there are four non-trivial diagrams for $\mathcal{P}^{\mathsf{MF}} \hat{V} \hat{\phi_{\v{q}}}$:
\begin{center}
\begin{tikzpicture}[scale=0.75]
\node (i) at (-1.5,0) {};
\node (f) at (2.5,0) {};
\node[dot,inner sep=1.5pt] (V1) at (0,0) {};
\node[dot,inner sep=1.5pt] (V2) at (0,1.5) {};
\node[draw,diamond,inner sep=1pt] (P) at (0,2.5) {$P$};
\node[draw,circle,inner sep=0.5pt] (phiq) at (1.5,0) {$\phi_q$};
\draw[postaction={on each segment={mid arrow}}] (f) -- (phiq) -- (V1) -- (i);
\draw[snake] (V1) -- (V2);
\draw[postaction={on each segment={mid arrow}}] 
(V2) -- ++(-0.3,0) to[bend left,out=90,in=90] (P.west)
(P.east) to[bend right,out=90,in=90]  ++(0.0,-1) -- (V2);
\end{tikzpicture}
\begin{tikzpicture}[scale=0.75]
\node (i) at (-1.5,0) {};
\node (f) at (1.5,0) {};
\node[dot,inner sep=1.5pt] (V1) at (0,0) {};
\node[dot,inner sep=1.5pt] (V2) at (0,1.5) {};
\node[draw,diamond,inner sep=1pt] (P) at (-0.5,2.5) {$P$};
\node[draw,circle,inner sep=0.5pt] (phiq) at (0.5,2.5) {$\phi_q$};
\draw[postaction={on each segment={mid arrow}}] (f) -- (V1)  -- (i);
\draw[snake] (V1) -- (V2);
\draw[postaction={on each segment={mid arrow}}] (V2) -- ++(-0.8,0) to[bend left,out=90,in=90] (P.west)
(P.east) -- (phiq.west)
(phiq.east) to[bend right,out=90,in=90]  ++(0.0,-1) -- (V2);
\end{tikzpicture}
\begin{tikzpicture}[scale=0.75]
\node (i) at (-2.75,0) {};
\node (f) at (2,0) {};
\node[draw,diamond,inner sep=1pt] (P) at (-1,0) {$P$};
\node[draw,circle,inner sep=0.5pt] (phiq) at (1,0) {$\phi_q$};
\node[dot,inner sep=1.5pt] (V1) at (-2,0) {};
\node[dot,inner sep=1.5pt] (V2) at (0,0) {};
\draw[snake] (0,0) arc (0:180:1);
\draw[postaction={on each segment={mid arrow}}]
(f) -- (phiq.east)
(phiq.west) -- (V2) -- (P.east)
(P.west) -- (V1) -- (i);
\end{tikzpicture}
\begin{tikzpicture}[scale=0.75]
\node (i) at (-2.75,0) {};
\node (f) at (2,0) {};
\node[draw,diamond,inner sep=1pt] (P) at (0.25,0) {$P$};
\node[draw,circle,inner sep=0.5pt] (phiq) at (-1.25,0) {$\phi_q$};
\node[dot,inner sep=1.5pt] (V1) at (-2,0) {};
\node[dot,inner sep=1.5pt] (V2) at (1,0) {};
\draw[snake] (1,0) arc (0:180:1.5);
\draw[postaction={on each segment={mid arrow}}]
(f) -- (V2) -- (P.east)
(P.west) -- (phiq.east)
(phiq.west) -- (V1) -- (i);
\end{tikzpicture}
\end{center}
The first two are Hartree-like, while the second two are Fock-like (and come with a minus sign). Since $\hat{\phi}_{\v{q}}$ acts before $\hat{V}$, there cannot be a $P$ separating $\phi_{\v{q}}$ from the interaction vertex. Comparing to the terms above, these diagrams gives
\begin{equation}
    \hat{h}_H[P] \hat{\phi}_{\v{q}}
    + \hat{h}_H[\phi_{\v{q}} P]
    + \hat{h}_F[P] \hat{\phi}_{\v{q}}
    + \hat{h}_F[\phi_{\v{q}} P].
\end{equation}
Similarly, $-\mathcal{P}^{\mathsf{MF}} \hat{\phi_{\v{q}}} \hat{V}$ gives
\begin{equation}
    - \hat{\phi}_{\v{q}} \hat{h}_H[P]
    - \hat{h}_H[P \phi_{\v{q}}]
    - \hat{\phi}_{\v{q}} \hat{h}_F[P]
    - \hat{h}_F[P \phi_{\v{q}}]. 
\end{equation}
Combining these
\begin{equation}
    \mathcal{P}^{\mathsf{MF}}[\hat{V},\hat{\phi_{\v{q}}}]
    = \big[\hat{h}_{H}[P] + \hat{h}_{F}[P], \hat{\phi}_{\v{q}}\big]
    + \hat{h}_{H}\big[ [\phi_{\v{q}}, P] \big] + \hat{h}_{F}\big[ [\phi_{\v{q}}, P]
    \big]
    = \big[\hat{h}_{H}[P] + \hat{h}_{F}[P], \hat{\phi}_{\v{q}}\big]
    - \hat{h}_{HF}\big[ [P,\phi_{\v{q}}]_{\v{q}} \big]
\end{equation}
The kinetic term $\mathcal{P}^{\mathsf{MF}}[\hat{h},\hat{\phi}_{\v{q}}] = [\hat{h},\hat{\phi}_{\v{q}}]$ combines with the first term here to give the full (non-boosted) Hartree-Fock Hamiltonian. Altogether then the action of $\L^{\textsf{MF}}_{\v{q}}$ on matrices $\phi_{\v{q}}$ is
\begin{equation}
    \L^{\textsf{MF}}_{\v{q}} \phi_q
    = \big[h_{HF}[P], \phi_{\v{q}}\big]_{\v{q}}
    - h_{HF}\big[ [P, \phi_{\v{q}}]_{\v{q}} \big].\label{eq:TDHF}
\end{equation}
The first term is the boosted commutator of the un-boosted HF Hamiltonian, and the second term is (minus) the boosted HF Hamiltonian of the boosted commutator.

\section{2PI Effective Action}
\label{app:2PI}
While we have focused on extracting the phonon action via Hartree-Fock in the main text, it is worth noting that our procedure is more general. In particular, any approximate scheme for computing the effective action $\Gamma[\v \theta]$ will do. The existence of such an effective action guarantees that computing for example the kineo-elastic coefficient by first shearing the unit cell, and then boosting, or boosting and the shearing will give the same result. Let us briefly discuss how (Time-dependent) Hartree-Fock can be considered such a scheme. In particular by considering the effective action of the Hamiltonian $H[\v \theta]$ as a function of the \emph{Green's function}, within the 2PI formalism \cite{BaymConvserving,Baym2PI}
\begin{equation}
   \Gamma[G,\v \theta]=\i\tr \log G^{-1}-\i \tr [GG_0^{-1}[\v \theta]-1]-\i\Phi[G,\v\theta]
\end{equation}
where, $G_0^{-1}[\v \theta]=\i \partial_t-h[\v \theta]$ is the free propagator and $\Phi$ is the Luttinger-Ward functional, which perturbatively is the sum of all 2 particle irreducible diagrams of the gauge twisted interaction. The effective action in terms of $\v \theta$ is obtained by finding the saddle point $G$ of $\Gamma$. In particular note that the equation of motion for $G$ is precisely the usual Dyson equations 
\begin{equation}
   0= \frac{\delta \Gamma}{\delta G}=\i G^{-1}-\i G_0^{-1}+\i\Sigma[G]
\label{eq:SD}
\end{equation}
where $\Sigma[G]=\frac{\delta \Phi[G]}{\delta G}$ is the self energy functional. 
Hartree-Fock emerges as the lowest order approximation to $\Phi$ in terms of interaction, which in this case is 
\begin{align} 
\Phi^\mathrm{HF}[G,{\v\theta}]&=\frac{1}{2}\int \dd \v r_1\dd \v r_2\,V(\v r_1+\v \theta(\v r_1,t)-\v r_2-\v\theta(\v r_2,t))G_{\alpha\alpha}(\v r_1,\v r_1)G_{\beta\beta}(\v r_2,\v r_2)\\ 
&- \frac{1}{2}\int \dd \v r_1\dd \v r_2\,V(\v r_1+\v \theta(\v r_1,t)-\v r_2-\v \theta(\v r_2,t))G_{\alpha\beta}(\v r_1,\v r_2)G_{\beta\alpha}(\v r_2,\v r_1)
\end{align} 
The equations of motion $\frac{\delta}{\delta G(x,y)}\log Z^\mathrm{HF}=0$ are solved precisely by the solutions to the self-consistent Hartree-Fock problem, with both stationary and time-dependent solutions corresponding to the conventional Hartree-Fock approach, and general time-dependent Hartree-Fock. Making the connection to the variational formulation can be done easily considering multiplying equation \eqref{eq:SD} with $G$, and taking the commutator:
\begin{equation}
\i[\partial_t,G]=[h+\Sigma^\mathrm{HF}[G],G]
\end{equation}
At the Hartree-Fock mean field level, $\Sigma$ becomes local in time, and we recover the most general time-dependent Hartree-Fock equations. By considering stationary solutions $[\partial_t,G]=0$, we recover the usual stationary Hartree-Fock equations $[h+\Sigma^\mathrm{HF}[G],G]=0$. By contrast the linearized TDHF we perform consist of taking a stationary solution $\bar{G}$, and considering small fluctuations $G=\bar{G}+\phi$. The linearized equations of motion for $\phi$ are 
\begin{equation}
    \i[\partial_t,\phi]=\left[h+\Sigma^\mathrm{HF}[\bar{G}],\phi\right]+\left[\frac{\delta\Sigma[\bar{G}]}{\delta G}\phi,\bar{G}\right]=\mathcal{L}[\phi]\label{eq:2PI_TDHF}
\end{equation}
which is precisely equation \eqref{eq:TDHF}. Suppose $U=e^{\i \alpha A}$ generates a continuous symmetry that commutes with the Hamiltonian, then $\phi_A=\i[A,\bar{G}]$ is an exact zero mode of equation \eqref{eq:2PI_TDHF}, which is a manifestation of the Goldstone theorem \cite{Goldstone}. Keeping higher order terms in $\Phi$ is a natural extension of HF.

\end{document}